\documentclass{elsart}
\usepackage{graphicx}
\usepackage{epsfig}
\newcounter{list1} 
\begin{document}

\begin{frontmatter}
\title{The continuous time random walk formalism in financial markets}
\author{Jaume Masoliver, Miquel Montero, Josep Perell\'o,}
\address{Departament de F\'{\i}sica Fonamental, 
Universitat de Barcelona, Diagonal, 647, 08028-Barcelona, Spain}
\author{and George H. Weiss}
\address{1105 N. Belgrade Rd. Silver Spring, Md. 20902, U.S.A.}

\date{\today}

\begin{abstract}
We adapt continuous time random walk (CTRW) formalism to describe asset price evolution and discuss some of the problems that can be treated using this approach. We basically focus on two aspects: (i) the derivation of the price distribution from high-frequency data, and (ii) the inverse problem, obtaining information on the market microstructure as reflected by high-frequency data knowing only the daily volatility. We apply the formalism to financial data to show that the CTRW offers alternative tools to deal with several complex issues of financial markets.
\end{abstract}

\begin{keyword}
continuous time random walk \sep volatility \sep financial markets \sep market microstructure
\end{keyword}
\end{frontmatter}

\section{Introduction}

The continuous time random walk (CTRW), first introduced by Montroll and Weiss (1965), has become a widely-used tool for studying the microstructure of random process appearing in a large variety of physical phenomena. These range from transport in disordered media (e.g., Montroll and Shlesinger 1984), earthquake modelling (e.g., Helmstetter and Sornette 2002) and even solar surface kinetics (e.g., Lawrence et al. 2001), to name just a few. This article extends the fields of application of the CTRW analysis by including the dynamics of financial markets. In this way, we aim to introduce the CTRW approach to the financial mathematics community where, to our knowledge, this formalism has been barely explored.

As its name suggests, the CTRW generalizes simple random walk models. Although the term 
``random walk" was coined by Pearson (1905), the formalism had been formulated in the seventeenth century in the context of gambling games such as the probability of ruin after betting $n$ times in a coin tossing game. In this case, the sum of gains minus losses in $n$ trials is equal to the state of a player's fortune (Weiss 1994). 

Financial markets have also been studied from the random walk point of view. In fact, this formalism was the first tentative model known in finance, having been suggested by Bachelier (1900) to describe stock market dynamics and give a price for a European call option. In fact, Bachelier modelled the price evolution assuming that prices change one unit at each time step with a probability $p$ of going up and $1-p$ of going down. Thus there are only two possible events. This process is called the binomial model and is the simplest random walk. Bachelier showed that the distribution of the resulting process, after a large number of time steps, tends to the Gaussian distribution. The limit behavior is often called the diffusion limit and the result obtained is just the consequence of the central limit theorem (Weiss). 

Several decades passed before there was further progress in the application of random walk methodology to analyze different aspects of financial transactions. Thus, in the book edited by Cootner (1964), there is a chapter devoted to the reexamination of the random walk hypothesis in finance. At that time, it was well-established that random walk models should be applied to the price return instead of the price itself as Bachelier asserted. In Cootner there are also several papers, basically published in the period 1961-1963, devoted to the question of whether the market is a random walk or a trend follower. Within these works we mention the article by Fama (1963) that studies an alternative to the Gaussian Bachelier random walk, proposing instead the Pareto distribution. 

Later on, Cox and Ross (1976) used the Bachelier ordinary random walk formalism to provide a discrete time analog to the well-known Black-Scholes option price. In addition, Cox and Ross also obtained a different limiting process when the number $n$ of time steps is large, finding that the binomial model leads to a Poisson jump process. Other contributions of the random walk approach to finance extend the binomial model by adding a third possible event, a crash, and observe the implications of it to the European option price (see e.g. Wilmott 1999). To our knowledge, there are not other financial market models exploiting all the possibilities that random walk analysis can offer in the study of many interesting phenomena in markets.

Note that the ordinary random walk formalism mentioned above is based on the assumption that step changes are made at equal time intervals but this is a first approximation for many physical phenomena and markets. The CTRW relaxes this restriction since it assumes that time intervals between transactions are not constant but random. Ticks have now, and in contrast with the Bachelier random walk, two sources of randomness: one coming from the amplitude and another one from the waiting times between ticks. This sophistication is necessary to describe markets with tick-by-tick data (i.e., the highest frequency price data available). The  deepest structure corresponds to the transaction-to-transaction operations. We believe that the CTRW formalism is especially useful in these cases and may improve the ordinary random walk approach. In contrast, it seems not to be suitable for cash indices since indices are averages of many prices and are recorded following some specific criteria described by an almost constant tick-by-tick frequency.

Despite this promising fact, the CTRW is hardly known among financial analysts. Physicists have recently provided, from a new discipline called econophysics, only a few examples of CTRW's applied to finance. Thus the papers by Scalas et al. (2000) and Mainardi et al. (2000) were among the first works addressed to this issue. Further developements were given by Raberto et al. (2002), Kutner and \'Swita{\l}a (2003), and Masoliver et al. (2003). Perhaps one of the most solid reasons in favor of CTRW models is that they provide general expressions for the distribution of prices at time $t$ in terms of two auxiliary densities that can be estimated from data: the probability density function (pdf) of the pausing time between ticks, $\psi(t)$, and the density for the magnitude of the price increment at a given tick, $h(x)$ (see Eqs.~(\ref{psi}) and~(\ref{h}) below for a formal definitions of these quantities). 

Other quantities, such as the distribution of daily or longer-time prices based on two probability density functions $\psi(t)$ and $h(x)$, can be obtained using the CTRW formalism. This in turn allows for the possibility of dealing with inverse problems, that is, estimating from the observed daily or longer-time data the forms of the microscopic functions $\psi(t)$ and $h(x)$. This is useful since in many practical situations, one only has, at most, daily data. In this case our formalism enables us to determine features of the otherwise unknown microscopic structure of the financial process.

Finally, we mention the existence of an important precedent that probably might have occurred to many readers. Within a very different approach, Merton (1976) proposed a jump market model whose jumps are completely independent and have a Poisson distribution for the waiting times. The CTRW formalism generalizes the Merton model because we have now the freedom of choosing any probability distribution for pausing times and jumps and, even more importantly, the present formalism allows one to study arbitrary correlations between them.

The paper is organized as follows. In Section \ref{fund} we present a short introduction to the CTRW formalism. In Section~\ref{general} we derive the exact distribution of prices and volatility in terms of market microstructure statistics, while in Section~\ref{inverse} we study the inverse problem. These sections contain analysis based on actual data to illustrate the procedures used. Finally, conclusions are drawn in Section~\ref{conclusions} and more technical questions are left to appendices available on the JEBO website.

\section{An introduction to the CTRW\label{fund}}

We first provide a definition of the CTRW from the perspective of financial markets. Let $Z(t)$ be the return defined by $Z(t)=\ln[S(t)/S(t_0)]$ where $S(t)$ is a speculative price and $t_0$ is an initial time. In what follows we will assume that the return process is stationary so that $Z(t)$ is independent of $t_0$. We are interested in the zero-mean return, $X(t)$,  rather than the return $Z(t)$. This is defined by
\begin{equation}
X(t)=Z(t)-\langle Z(t)\rangle,
\label{x}
\end{equation}
where $\langle Z(t)\rangle$ is the average of $Z(t)$. 

\begin{figure}
\begin{center}
\epsfig{file=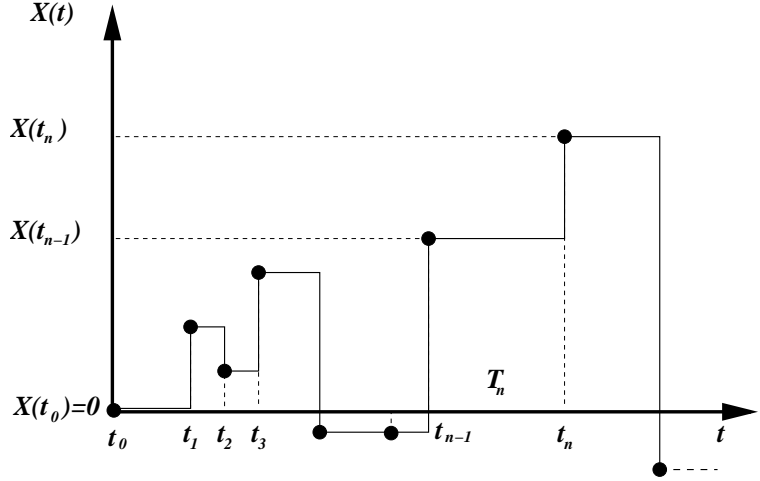,width=10cm}
\end{center}
\caption{Schematic representation of the return process. The dots mark the value $X(t_n)$ of the return after each sojourn. $T_n=t_n-t_{n-1}$ is the time increment of the $n$-th sojourn.}
\label{fig1}
\end{figure}

We now suppose that $X(t)$ evolves following a CTRW and that any realization consists of a series of step functions (see Fig. \ref{fig1}). Therefore the return evolves discontinuously, and during any sojourn its value remains constant. In this picture $X(t)$ changes at random times $t_0,t_1, t_2,\cdots,t_n,\cdots$. We assume that the intervals between successive steps, which we call sojourns or pausing times, $T_n=t_n-t_{n-1}$ 
($n=1,2,3,\cdots$) are independent, identically distributed, random variables with a probability density function, $\psi(t)$, defined by 
\begin{equation}
\psi(t)dt={\rm Prob}\{t<T_n\leq t+dt\}.
\label{psi}
\end{equation}

At the conclusion of a given sojourn $X(t)$ undergoes a random change equal to $\Delta X_n=X(t_n)-X(t_{n-1})$ whose probability density function is defined by
\begin{equation}
h(x)dx={\rm Prob}\{x<\Delta X_n\leq x+dx\}.
\label{h}
\end{equation}
We collect these two random sources into one single function $\rho(x,t)$ that represents the joint probability density function of pausing times and return increments. That is,
\begin{equation}
\rho(x,t)dx dt={\rm Prob}\{x<\Delta X\leq x+dx;\, t<T\leq t+dt\}.
\label{rhodef}
\end{equation}
In what follows, we take $\rho(x,t)$ to be an even function of $x$ thus assuring that there is no net drift in the evolution of $X(t)$. Observe that two marginal densities can be formed out of $\rho(x,t)$; the pausing-time density $\psi(t)$ and the density for a random jump $h(x)$, these are related to $\rho(x,t)$ by
\begin{equation}
\psi(t)=\int_{-\infty}^{\infty}\rho(x,t)dx,\qquad 
h(x)=\int_{0}^{\infty}\rho(x,t)dt.
\label{psib}
\end{equation}

Our main goal is to obtain the probability density function of $X(t)$. This function is the propagator, defined by
$$
p(x,t)dx={\rm Prob}\{x<X(t)\leq x+dx\}.
$$
It is well known that $p(x,t)$, satisfies the following renewal equation (e.g., Cox 1965 and Weiss 1994),
\begin{equation}
p(x,t)=p_0(x,t)+\int_0^tdt'\int_{-\infty}^{\infty}\rho(x',t')p(x-x',t-t')dx',
\label{inteq}
\end{equation}
where we have assumed that the initial jump occurred at $t=0$. Equation (\ref{inteq})  involves several new definitions. The function $p_0(x,t)$ is the propagator prior to the first jump and is equal to 
\begin{equation}
p_0(x,t)=\Psi(t)\delta(x),
\label{p0}
\end{equation}
where $\Psi(t)$ is the probability that no transaction has occurred before time $t$,
\begin{equation}
\Psi(t)=\int_t^\infty \psi(t') dt'.
\label{Psi}
\end{equation}
The function $\delta(x)$ appearing in Eq.~(\ref{p0}) is the Dirac delta function that can be defined by $\delta(x)=0$ for $x\neq 0$ and 
$$
\int_{-\infty}^{\infty}\delta(x)f(x)dx=f(0)
$$
for any integrable function $f(x)$ (Lighthill 1980). Equation (\ref{inteq}) is derived from the consideration that at time $t$ the first transaction has not yet occurred, this given by the first term on the right hand side of this equation, or else that a transaction occurred at time $t'<t$, at which time the return had value $x'$ and from $(x',t')$ the return process is renewed. We can solve Eq.~(\ref{inteq}) in terms of the joint Fourier-Laplace transform:
$$
\hat{p}(\omega,s)=
\int_0^\infty dte^{-st}\int_{-\infty}^{\infty}dx e^{i\omega x}p(x,t).
$$
The solution is
\begin{equation}
\hat{p}(\omega,s)=\frac{\hat{p}_0(\omega,s)}{1-\hat{\rho}(\omega,s)},
\label{hatp}
\end{equation}
where $\hat{p}_0(\omega,s)$ and $\hat{\rho}(\omega,s)$ are respectively the joint Fourier-Laplace transforms of the functions $p_0(x,t)$ and $\rho(x,t)$. We easily see from 
Eq. (\ref{p0}) that the explicit form of $\hat{p}_0(\omega,s)$ is given by
\begin{equation}
\hat{p}_0(\omega,s)=\frac{1-\hat{\psi}(s)}{s},
\label{hatp0}
\end{equation}
where $\hat{\psi}(s)$ is the Laplace transform of the pausing-time density $\psi(t)$. Equation~(\ref{hatp}) furnishes the complete solution in the transform domain and is a convenient starting point in any CTRW analysis.

\begin{figure}[hptb]
\begin{tabular}{l}
(a)\\
\epsfig{file=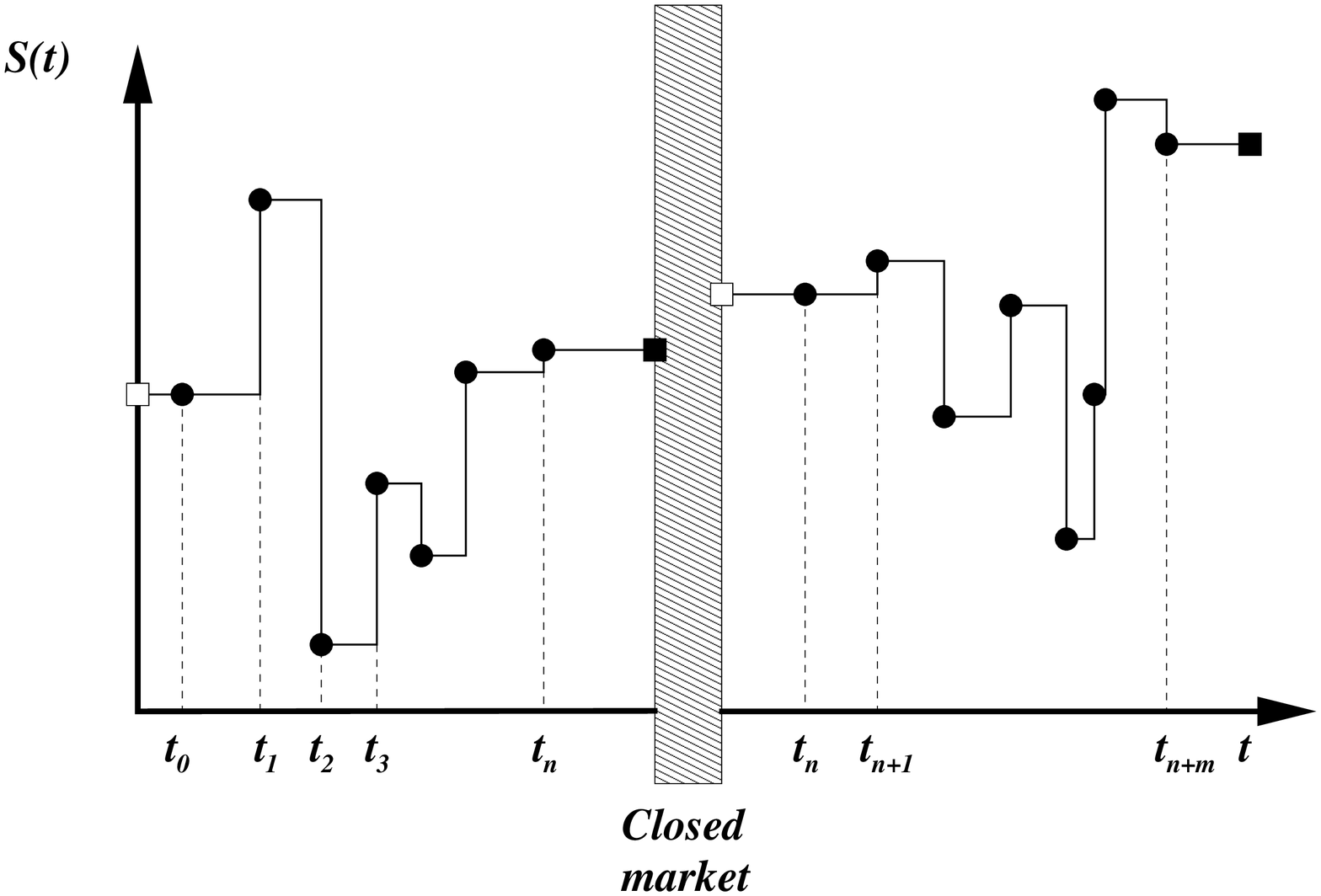,width=10cm}
\\
(b)\\
\epsfig{file=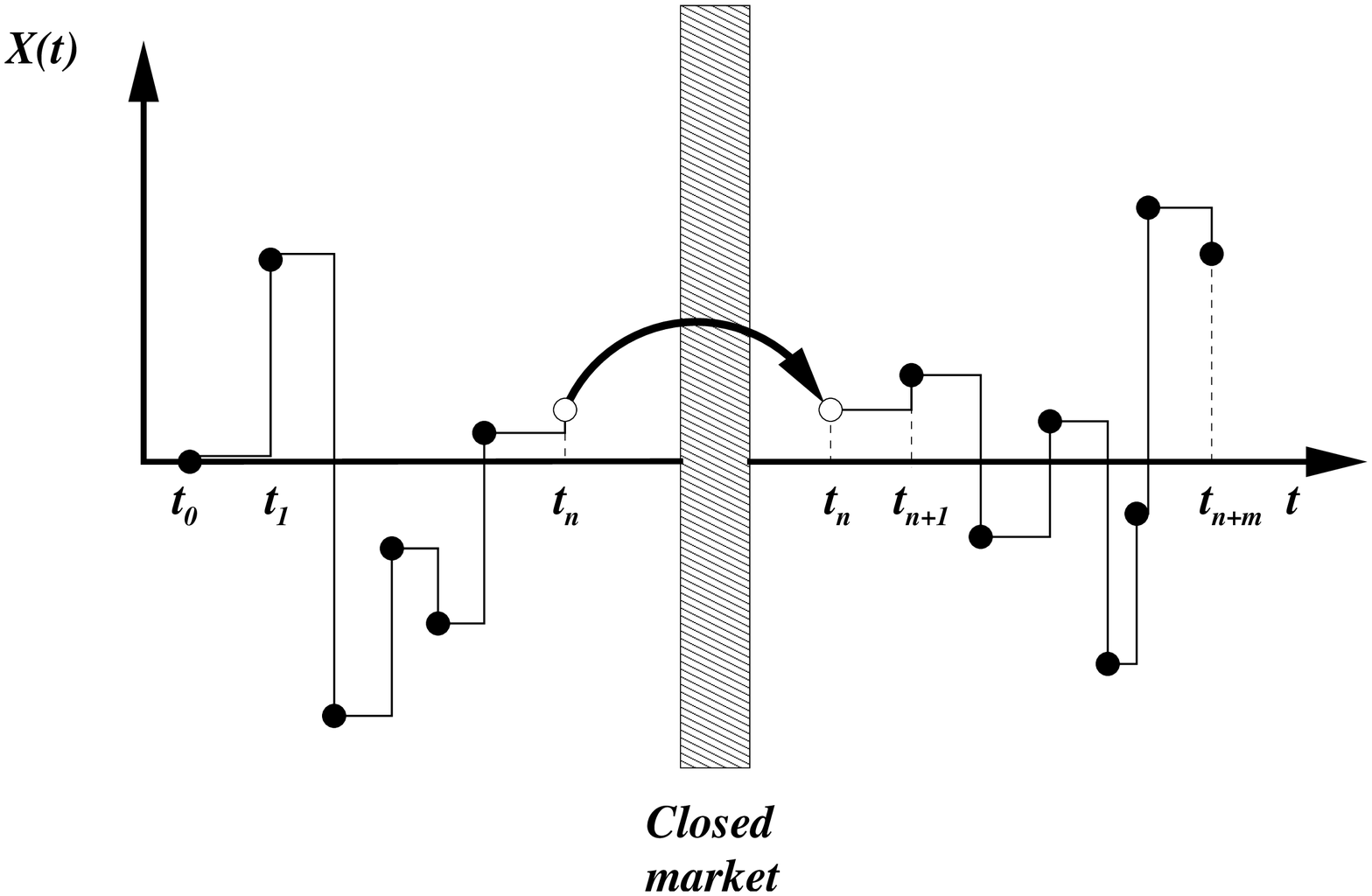,width=10cm}
\end{tabular}
\caption{In (a) we describe the evolution of the stock price $S(t)$. Consecutive transactions take place at discrete random times $t_0,t_1,t_2,\cdots$ and are represented by black circles. At every moment, the last traded price is the reference value during the intervals between changes. Two prices are noteworthy: the {\it opening} and {\it closing} prices. These are respectively the price of the first and last transaction and are represented by white (opening) and black (closing) boxes. In many situations opening and closing prices are the only data available. Generally the closing price of a session does not coincide with the opening price of the next day. Nevertheless, as an approximation, the last return of a given day can be taken as the first return of the next day. Therefore, since we do not model the behavior of the market when it is closed, we proceed as shown in (b); that is, we identify the last operation of a given day with the first transaction of the next day. In the figure both transactions are represented by white circles.}
\label{fig1b}
\end{figure}

One may argue that the CTRW is only applicable to intraday price evolution. The process can be extended to larger time lags as depicted in Fig.~\ref{fig1b}. Neglecting overnight effects on the price, obviously a first approximation, we show there how to describe the daily price evolution and thus represent the whole historical price evolution within the framework of the CTRW. This allows that the time $t$ appearing in $p(x,t)$ can be any time interval: minutes, hours, days, or longer. A significant effect of this linkage is that we unify within the same formalism high-frequency data with low-frequency data such as daily or weekly closing prices. We will see in future sections that this enables us to derive some important results.

\section{From market microstructure to return distributions\label{general}}

We want to obtain the return distribution from information concerning tick-by-tick price behavior. Apparently, this is straightforward by taking the Fourier-Laplace inversion of Eq.~(\ref{hatp}). Unfortunately, we need first to determine the form of the joint density $\rho(x,t)$, but this is not possible from the available data. More easily accessible are the marginal pdf's, $h(x)$ and $\psi(t)$, that fully define the tick-by-tick statistics. It is therefore essential to assume a functional relation between $\rho(x,t)$ and its marginal densities $\psi(t)$ and $h(x)$. Let us first study some possible ways of combining $\psi(t)$ and $h(x)$ to obtain $\rho(x,t)$ and discuss their main implications. 

The simplest choice would be based on the assumption that return increments and their duration time are independent random variables (Scalas et al.). In this case
({\it ansatz 1})
\begin{equation}
\rho(x,t)=h(x)\psi(t),
\label{independent}
\end{equation}
so that Eq. (\ref{hatp}) becomes

\begin{equation}
\hat{p}(\omega,s)=\frac{[1-\hat{\psi}(s)]/s}
{1-\tilde{h}(\omega)\hat{\psi}(s)}.
\label{independent2}
\end{equation}
For the case of exponentially distributed pausing times, $\psi(t)=\lambda \exp(-\lambda t)$, and after Laplace inversion, equation (\ref{independent2}) results in the following characteristic function:
\begin{equation}
\tilde{p}(\omega,t)=\exp\{-\lambda[1-\tilde{h}(\omega)]t\}.
\label{independent3}
\end{equation}

However, in many situations one expects some degree of correlation between return increments and their duration, while Eq. (\ref{independent}) implies complete independence between them. This is, for example, confirmed in the statistical analysis by Raberto et al. for the specific stock shares of  General Electric. 

One possible assumption is to suppose that the density $\rho(x,t)$ is such that its characteristic function $\tilde{\rho}(\omega,t)$ has the following functional form ({\it ansatz 2}),
\begin{equation} 
\tilde{\rho}(\omega,t)=\psi\left[\frac{t}{\tilde{h}(\omega)}\right],
\label{fform}
\end{equation}
where $\psi(t)$ is the pausing time density and $\tilde{h}(\omega)$ is the Fourier transform of $h(x)$. The assumption given by Eq. (\ref{fform}) is at least intuitively plausible because it implies that one must wait for a long time in order for a large variation of return to occur. In other words, large increments of the return are very infrequent. 

Let us show this by proving that sojourn times, $T=t_n-t_{n-1}$, and return quadratic increments, $\Delta X^2=[X(t_n)-X(t_{n-1})]^2$, are positively correlated so that  increasing return variations imply increasing sojourn times and vice versa. In effect, we define the following measure of the correlation between $\Delta X^2$ and $T$ by
\begin{equation}
r=\frac{\langle\Delta X^2 T\rangle}{\langle\Delta X^2\rangle\langle T\rangle}-1.
\label{r}
\end{equation}
We can easily evaluate the cross average $\langle\Delta X^2 T\rangle$ using the joint characteristic function $\tilde{\rho}(\omega,t)$.Thus
$$
\langle\Delta X^2 T\rangle=-\left.\frac{\partial^2}{\partial\omega^2}
\int_0^\infty t\tilde{\rho}(\omega,t)dt\right|_{\omega=0},
$$
which after using Eq. (\ref{fform}) yields 
$\langle\Delta X^2 T\rangle=2\langle\Delta X^2\rangle\langle T\rangle.$ Hence $r=1$ as we meant to prove.

\begin{figure}
\centerline{\includegraphics[width=14cm]{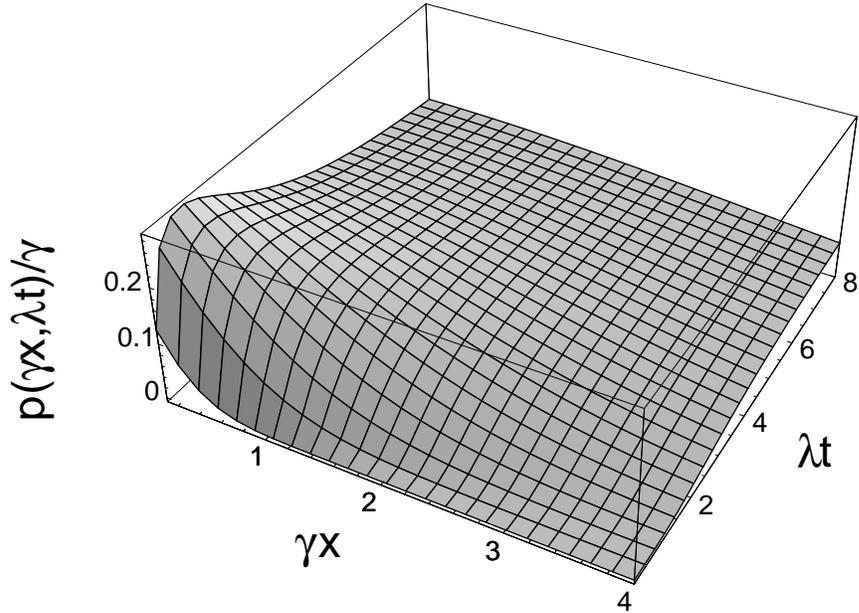}}
\caption{The return pdf in dimensionless units $p(\gamma x,\lambda t)/\gamma$  given in Eq. (\ref{poissonlaplace}) for $x>0$, that is, without the delta function term appearing in~(\ref{poissonlaplace}). By the assumed symmetry in $x$, $p(x,t)$ behaves similarly for negative returns.} 
\label{fig1c}
\end{figure}

We now come back to the ansatz in Eq.~(\ref{fform}) and observe that it allows us to write $\hat{\rho}(\omega,s)$ in the form $\hat{\rho}(\omega,s)=\tilde{h}(\omega)\hat{\psi}[s\tilde{h}(\omega)]$. This in turn leads us to write the formal solution to the problem given by Eq. (\ref{hatp}) in the following more explicit form:
\begin{equation}
\hat{p}(\omega,s)=\frac{[1-\hat{\psi}(s)]/s}
{1-\tilde{h}(\omega)\hat{\psi}\left[s\tilde{h}(\omega)\right]}.
\label{formalsolution}
\end{equation}
We next consider an example in which it is possible to evaluate $p(x,t)$ directly. Thus, on the assumption that $\psi(t)=\lambda \exp(-\lambda t)$ and $h(x)=\gamma \exp(-\gamma|x|)/2$, Eq.~(\ref{formalsolution}) yields
\begin{equation}
\hat{p}(\omega,s)=
\frac{1}{\lambda+s}\left[1+\frac{\lambda}{\lambda\omega^2/\gamma^2+s}\right]. 
\label{poissoncase3}
\end{equation}
The inverse Fourier transform followed by the Laplace inversion and the convolution theorem lead us to the following return pdf:
\begin{equation}
p(x,t)=e^{-\lambda t}\left[\delta(x)+\frac{\gamma}{\sqrt{\pi}}
\int_0^{\sqrt{\lambda t}}e^{\xi^2-\gamma^2x^2/4\xi^2}d\xi\right].
\label{poissonlaplace}
\end{equation}
In Fig. \ref{fig1c} we plot this density for positive returns $x>0$.

\begin{figure}
\centerline{\includegraphics[width=13cm]{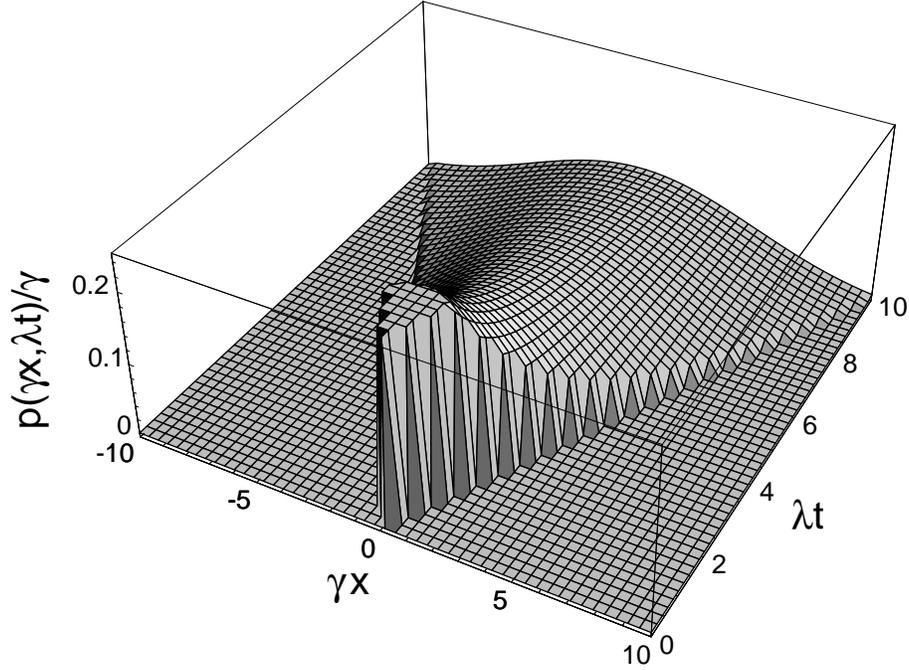}}
\caption{The return pdf in dimensionless units $p(\gamma x,\lambda t)/\gamma$  given in Eq. (\ref{poissonlaplace2}) for $x\neq 0$, without the delta function term.}
\label{fig1cc}
\end{figure}

The ansatz given in Eq. (\ref{fform}) is not the only possible one that assumes a dependence between jumps and pausing times. Another plausible choice would be given by the following form ({\it ansatz 3}):
\begin{equation}
\hat{\rho}(\omega,s)=\hat{\psi}(s)\tilde{h}[\omega\hat{\psi}(s)].
\label{hatrho2}
\end{equation}
Following the method outlined above one can easily see that in this case return variations and pausing  times are also positively correlated but with a higher correlation (cf. Eq.~(\ref{r})) given by $r=2$.

Note that in this case return pdf is given by
\begin{equation}
\hat{p}(\omega,s)=\frac{1}{\lambda+s}+
\frac{\lambda}{s(\lambda+s)+\lambda^2\omega^2/\gamma^2} 
\label{poissoncase4}
\end{equation}
and its joint Fourier-Laplace inversion yields
\begin{equation}
p(x,t)=e^{-\lambda t}\delta(x)+(\gamma/2)e^{-\lambda t/2}\Theta(\lambda t-\gamma|x|)
I_0\left(\sqrt{\lambda^2t^2-\gamma^2x^2}/2\right),
\label{poissonlaplace2}
\end{equation}
where $\Theta(x)$ is the Heaviside step function defined by
$$
\Theta(x)=\left\{
\begin{array}{ll}
1 & \qquad \mbox{when } x>0,\\ 
0 & \qquad \mbox{when } x<0;
\end{array}
\right.
$$
and $I_0(x)$ is the modified Bessel function. We note that, contrary to the pdf's given in Eqs. (\ref{independent3}) and~(\ref{poissonlaplace}), the density $p(x,t)$ given by Eq. (\ref{poissonlaplace2}) has a compact support since the value of the zero-mean return, $X(t)$, differs from zero only within the interval $-\lambda t/\gamma<x<\lambda t/\gamma$ for any fixed time $t$. Since this limitation on the value of $X(t)$ is unrealistic, the model given by the ansatz in (\ref{hatrho2}) together with the assumption that jumps are exponentially distributed seems not to describe events in real financial markets. In any case, for completeness, we plot this pdf in Fig. \ref{fig1cc}.

\subsection{Some asymptotic results}\label{sec:asymp}

We now focus on some asymptotic expressions of the distribution of returns, exemplified by its characteristic function $\tilde{p}(\omega,t)$. These approximate expressions are completely general and independent of the model chosen for $\rho(x,t)$ as long as the sojourn time density $\psi(t)$ has some finite moments. Under these conditions, we have derived in Appendix A (see also Kotulski 1995 for an alternative method) some interesting results that can be summarized as follows:
\begin{list}
{(\alph{list1})}{\usecounter{list1}}
\item 
If the jumps pdf $h(x)$ has a finite second moment, $\mu_2=\langle \Delta X^2\rangle<\infty$, the asymptotic distribution of returns for long times approaches to the Gaussian density:
\begin{equation}
\tilde{p}(\omega,t)\simeq e^{-\mu_2\omega^2t/2\langle T\rangle} \qquad (t\gg\langle T\rangle).
\label{gaussian}
\end{equation}

\item 
If $h(x)$ is a long-tailed density, $h(x)\sim|x|^{-1-\alpha}$ as $|x|\rightarrow\infty$, then $\tilde{h}(\omega)$ has an expansion of the form  $\tilde{h}(\omega)\simeq 1-k|\omega|^{\alpha}$ as $\omega\rightarrow 0$ (note that if $0<\alpha<2$, $h(x)$ has infinite variance). Moreover, if we assume that for $\omega$ small $\langle Te^{i\omega\Delta X}\rangle\simeq\langle T\rangle$, then the asymptotic return pdf approaches the L\'evy distribution:
\begin{equation}
\tilde{p}(\omega,t)\simeq e^{-k|\omega|^\alpha t/\langle T\rangle} \qquad (t\gg\langle T\rangle).
\label{levy}
\end{equation}

\item
At intermediate times, $t\approx\langle T\rangle$, the behavior of $p(x,t)$ for large values of $|x|$ is the same as that of the jump distribution: 
\begin{equation}
p(x,t)\sim \frac{t}{\langle T\rangle}h(x).
\label{tailsp}
\end{equation}
\end{list}
Note incidentally that the L\'{e}vy distribution has been proposed by several authors as a good candidate for describing financial distributions when the observed fat tails are not Gaussian (e.g., Mandelbrot 1963, Fama 1963). However, from the discussion above, we see that in the framework of CTRW's the appearance of L\'{e}vy distributions is linked to the existence of transactions for which the jumps have infinite variance. Since this is manifestly impossible in real markets, we conclude that the L\'{e}vy distribution is an unsuitable candidate for the propagator. In order to overcome this inconsistency some authors propose a truncated L\'evy distribution (see, for instance, Mantegna and Stanley 1995), although in many cases the truncation procedure is done in an arbitrary way, without any justification provided by the behavior of the market.

\subsection{The volatility}

Aside from the pdf $p(x,t)$, which provides maximal information about the evolution of $X(t)$, there is another quantity of considerable practical interest: the volatility, which does not require the knowledge of the entire jump distribution $h(x)$. It suffices to know the pdf $\psi(t)$ and the first two moments of $h(x)$.

Let us denote by $\hat{m}_2(s)$ the Laplace transform of the second moment $\langle X^2(t)\rangle$:
$$
\hat{m}_2(s)\equiv\int_0^\infty e^{-st}\langle X^2(t)\rangle dt.
$$
This quantity can be derived in terms of the joint Fourier-Laplace transform of $p(x,t)$ by
\begin{equation}
\hat{m}_2(s)=
-\left.\frac{\partial^2\hat{p}(\omega,s)}{\partial\omega^2}\right|_{\omega=0}.
\label{moments1}
\end{equation}
The combination of Eqs. (\ref{hatp}) and (\ref{moments1}) leads to the relation
\begin{equation}
\hat{m}_2(s)=\frac{\hat{R}_2(s)}{s[1-\hat{\psi}(s)]},
\label{volatilitygeneral}
\end{equation}
where $\hat{R}_2(s)=-\left.\partial^2\hat{\rho}(\omega,s)/\partial\omega^2\right|_{\omega=0}$.

For the independent model defined by Eq. (\ref{independent}) we see that 
$\hat{R}_2(s)=\mu_2\hat{\psi}(s)$, where $\mu_2=\langle\Delta X^2\rangle$ is the second moment of the jump density. Hence
\begin{equation}
\hat{m}_2(s)=\mu_2\frac{\hat{\psi}(s)}{s[1-\hat{\psi}(s)]}.
\label{volatilityindependent}
\end{equation}
Note that the assumption of exponentially distributed sojourns, for which
$\hat{\psi}(s)=\lambda/(\lambda+s)$, leads to 
\begin{equation}
\langle X^2(t)\rangle=\lambda\mu_2t,
\label{poissonindependent}
\end{equation}
so that the variance always increases linearly with time as in the Wiener process. 

For the model specified by the ansatz (\ref{fform}) we have $\hat{R}_2(s)=\mu_2[s\hat{\psi}(s)]'$, where the prime denotes a derivative, and
\begin{equation}
\hat{m}_2(s)=\frac{\mu_2[s\hat{\psi}(s)]'}
{s[1-\hat{\psi}(s)]}.
\label{volatility}
\end{equation}
For exponentially distributed pausing times we have
$\hat{m}_2(s)=\mu_2\lambda^2/s^2(\lambda+s)$, so that
\begin{equation}
\langle X^2(t)\rangle=\mu_2\left[\lambda t+e^{-\lambda t}-1\right].
\label{volatilitypoisson}
\end{equation}
The limiting behaviors of this volatility are $\langle X^2(t)\rangle\sim\mu_2(\lambda t)^2$ for $\lambda t\ll 1$, and $
\langle X^2(t)\rangle\sim\mu_2\lambda t$ for $\lambda t\gg 1$. 

For the second ansatz given by Eq.~(\ref{hatrho2}) we see that $\hat{R}_2(s)=\mu_2\hat{\psi}(s)^3$ and 
\begin{equation}
\hat{m}_2(s)=\frac{\mu_2\hat{\psi}(s)^3}
{s[1-\hat{\psi}(s)]}.
\label{volatility2}
\end{equation}
In the case of exponentially distributed pausing times we have 
$$
\hat{m}_2(s)=\lambda^3\mu_2/s^2(\lambda+s)^2
$$
and 
\begin{equation}
\langle X^2(t)\rangle=\mu_2\left[\lambda t\left(1+e^{-\lambda t}\right)-2\left(1-e^{-\lambda t}\right)\right].
\label{volatilitypoisson2}
\end{equation}
In this case the limiting behaviors are given by $\langle X^2(t)\rangle\sim\mu_2(\lambda t)^3/6$ for $\lambda t\ll 1$  and $\langle X^2(t)\rangle\sim\mu_2\lambda t$ for $\lambda t\gg 1$. Therefore, and similar to ansatz (\ref{fform}), if we suppose that sojourn times are exponentially distributed, the volatility has a superdiffusion-like behavior at short times and a diffusion-like behavior at long times. 

We remark that the asymptotic variance grows linearly with time regardless of the specific form of $\rho(x,t)$. Indeed, assuming that in all realistic models $\mu_2$ and $\langle T\rangle$ are finite  from Eq.~(\ref{gaussian}) we readily obtain 
\begin{equation}
\langle X^2(t)\rangle\simeq \frac{\mu_2}{\langle T\rangle}\ t \qquad (t\gg \langle T\rangle),
\label{asympvol}
\end{equation}
independent of the correlation model.

\subsection{The U.S. dollar/Deutsche mark case}
\label{rp}

Let us now show how to infer the return statistics at an arbitrary time $t$ from high frequency data. The case we discuss is that of the U.S. dollar/Deutsche mark future exchange, based on a transactions data base from January 1993 to December 1997. This consists of 1,048,590 transactions. We have chosen this data since this future has a high trading volume that is required to provide reliable statistical inferences.

The first assignment is to estimate from high-frequency data plausible forms for $\psi(t)$ and $h(x)$. In Fig. \ref{fig2} we plot the experimental pausing-time density $\psi(t)$ for the U.S. dollar/Deutsche mark future. We see that the data is well fit by a power-law density:
\begin{equation}
\psi(t)=\frac{\lambda(\alpha-1)}{(1+\lambda t)^\alpha},
\label{psifit}
\end{equation}
where $\alpha=3.47$ and $\lambda=2.73\times 10^{-2} s^{-1}$. The mean sojourn time and its second moment are
\begin{equation}
\langle T\rangle=\frac{\lambda^{-1}}{\alpha-2},\qquad
\langle T^2\rangle=\frac{2\lambda^{-2}}{(\alpha-2)(\alpha-3)}.
\label{meantime}
\end{equation}
For the dollar/mark future market the experimental mean sojourn time directly evaluated from data is $\langle T\rangle_{exp}=23.6\ s$, in satisfactory agreement with the theoretical prediction of $\langle T\rangle=24.9\ s$ evaluated from Eq.~({\ref{meantime}) with the parameters estimated from the fit to $\psi(t)$ in Fig.~\ref{fig2}.

\begin{figure}
\begin{center}
\epsfig{file=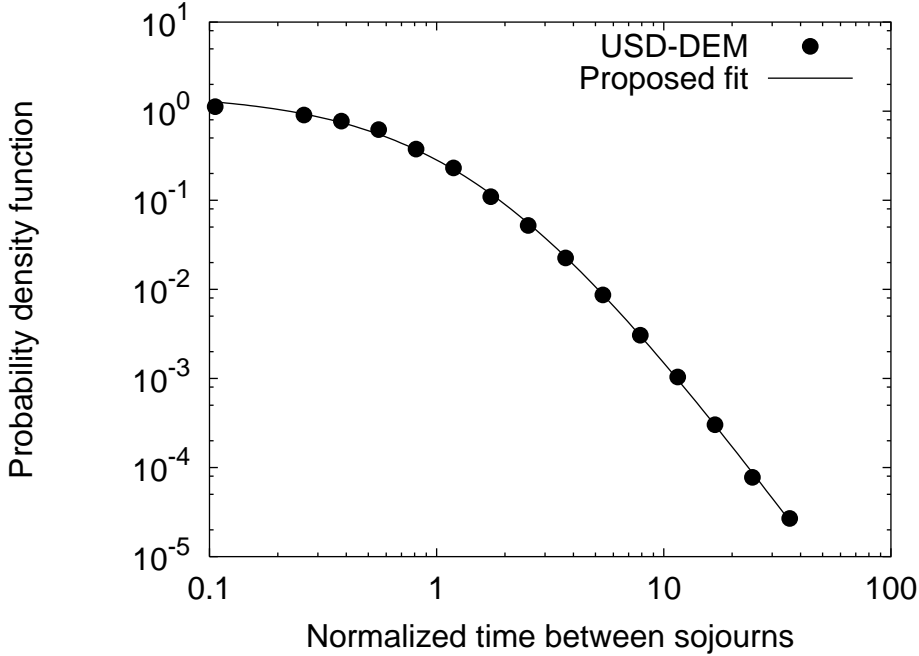}
\end{center}
\caption{Empirical distribution of the time between the closest-to-maturity transactions of the U.S. dollar/Deutsche mark future (in the US market). The analyzed data range from January, 1993, to December, 1997. The pdf of the sojourn times, $\psi(t)$, clearly follows a power law at sufficiently long times. The solid line is the representation of Eq. (\ref{psifit}). The normalized time between sojourns is $t/\langle T\rangle$.} 
\label{fig2}
\end{figure}

\begin{figure}
\begin{center}
\epsfig{file=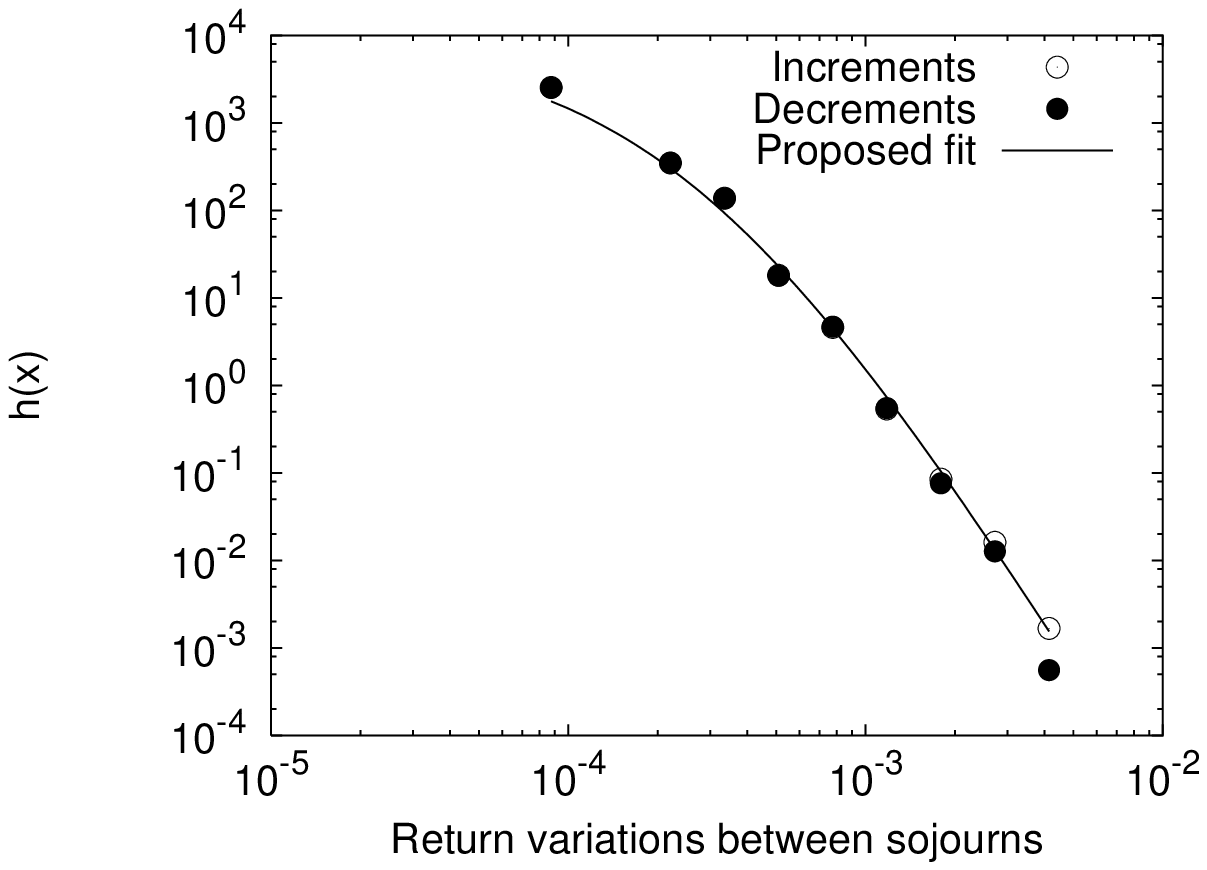}
\end{center}
\caption{Empirical distribution of the logarithmic changes between transactions in the U.S. dollar/Deutsche mark future market. Positive variations (increments) and negative variations (decrements) exhibit approximately the same behavior, thereby supporting our assumption of the symmetry of $h(x)$. The plot suggests a power law dependence at large return increments. It also includes a graph showing the shape of $h(x)$ in Eq.~(\ref{hfit}), using the parameters reported there.} 
\label{fig3}
\end{figure}

In Fig. \ref{fig3} we plot the experimental jump density, $h(x)$. The data indicate that the experimental $h(x)$ can be considered a symmetric function of the return increment $x$. A good fit is also given by a power law with a greater exponent than that of $\psi(t)$:
\begin{equation}
h(x)=\frac{(\beta-1)}{2\gamma(1+|x|/\gamma)^{\beta}},
\label{hfit}
\end{equation}
where $\beta=5.52$ and $\gamma=2.64\times 10^{-4}$. Power law densities such as (\ref{psifit}) and (\ref{hfit}) have been suggested for describing several market models such as individual companies by Plerou et al. (1999), market indices by Gopikrishnan et al. (1999) and Mainardi et al. (2000).

We have now all the elements to get the return pdf $p(x,t)$ at any time $t$ and, hence, to obtain a complete analysis of the dollar/mark exchange market. In order to focus ideas and results, we will now assume that pausing times and jump sizes are positively correlated as specified in Eq. (\ref{fform}). This is supported by empirical data since for the U.S. dollar/Deutsche mark future market we have a positive correlation given by $r=0.53$ (cf. Eq. (\ref{r})).

Unfortunately, the densities given by Eqs.~(\ref{psifit}) and (\ref{hfit}) do not allow an  exact evaluation of $p(x,t)$, or even of the characteristic function $\tilde{p}(\omega,t)$. The Fourier and Laplace transforms of $h(x)$ and $\psi(t)$ are too complex to manage them analytically as required for deriving the characteristic function~(\ref{formalsolution}). We will therefore restrict ourselves to study the behavior of the tails of the distribution. We have shown above that, for moderate times, the tails of $p(x,t)$ are the same as that of the jump density $h(x)$. That is, for large $|x|$ (cf. Eq.~(\ref{tailsp})),
$$
p(x,t)\sim\frac{t}{\langle T\rangle} h(x) \qquad\left(t\approx\langle T\rangle\right).
$$
In our case we see from Eqs.~(\ref{meantime})-(\ref{hfit}) that $\langle T\rangle=\lambda^{-1}/(\alpha-2)$ and $h(x)\sim(\beta-1)\gamma^{\beta-1}|x|^{-\beta}/2$ as $|x|\rightarrow\infty$. Therefore,
\begin{equation}
p(x,t)\sim\frac{(\beta-1)\lambda t}{2(\alpha-2)}\ \frac{\gamma^{\beta-1}}{|x|^{\beta}} 
\qquad\left(t\approx\langle T\rangle\right).
\label{tailp}
\end{equation}
Hence, the tails of $p(x,t)$ follow the same power law as that of the return increment distribution $h(x)$. Observe that the approximation~(\ref{tailp}) is valid when the time lag is of the same order of  the mean sojourn. This prediction of the theoretical model is confirmed by actual data. In Fig. \ref{fig5} we show the empirical $p(x,t)$ for the dollar/mark future exchange for $t=15$ seconds in which the time lag $t$ is of the same order than  the empirical mean sojourn $\langle T\rangle_{exp}=23.9\ s$. The empirical distribution clearly shows a power-law decay with exponent $\beta\approx 5.5$, which coincides with the decaying exponent of $h(x)$ which is consistent with the predictions of the CTRW model. 

We close this section with a remark about the volatility. In this case, we need to obtain the Laplace transform of the expression for $\psi(t)$ given by Eq.~(\ref{psifit}). This results in an expression in terms of the Kummer function $U(a,c,z)$ (Magnus et al. 1966), which renders very difficult the inversion of $\hat{m}_2(s)$ to obtain $\langle X^2(t)\rangle$. We thus limit ourselves to another asymptotic result. As we have proven above, the variance grows linearly with $t$ as $t\rightarrow\infty$ regardless of the model chosen; therefore, 
$\langle X^2(t)\rangle\sim\mu_2t/\langle T\rangle$ (see Eq. (\ref{asympvol})). In our case 
$\langle T\rangle=\lambda^{-1}/(\alpha-2)$ and $\langle X^2(t)\rangle\simeq(\alpha-2)\mu_2\lambda t$ for 
$\lambda t\gg 1$. We refer the reader to Fig.~\ref{fig10} of Sect. \ref{marketapplication} where this asymptotic linear dependence on time is clearly observed for the U.S. dollar/Deutsche mark futures market.

\begin{figure}
\begin{center}
\epsfig{file=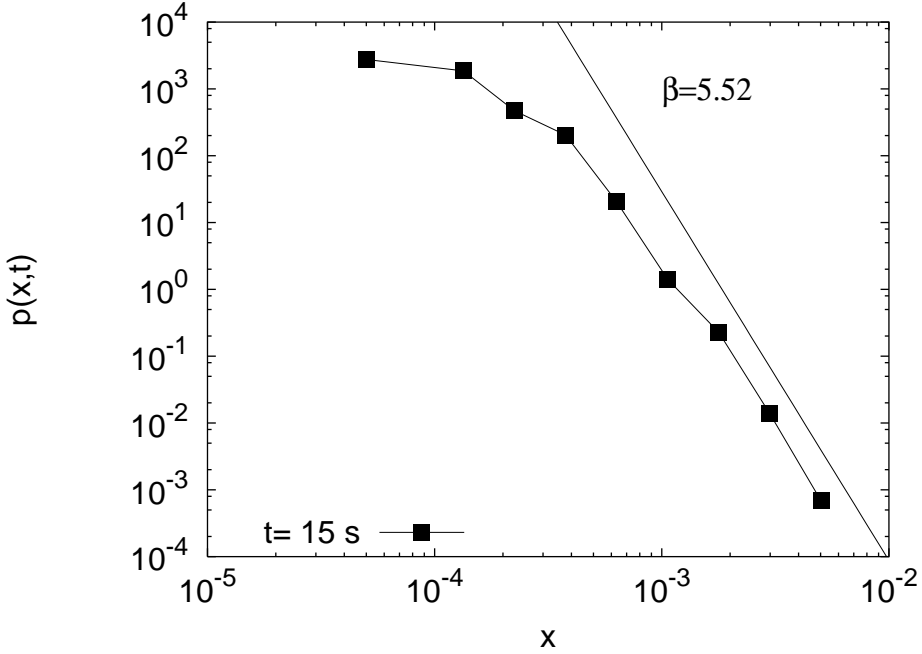}
\end{center}
\caption{The empirical probability density function $p(x,t)$ for a time lag $t$ of 15 seconds (U.S. dollar/Deutsche mark future market). The model leads to a power-law decay, governed by an exponent $\beta$ that is precisely the one appearing in the power-law for the jump density $h(x)$ as expected when $t\approx\langle T\rangle_{exp}=23.9\ s$. We show the tail only for positive increments; the tail for negative increments behaves similarly.}
\label{fig5}
\end{figure}

\section{The inverse problem\label{inverse}}

\begin{figure}
\begin{tabular}{cc}
\epsfig{file=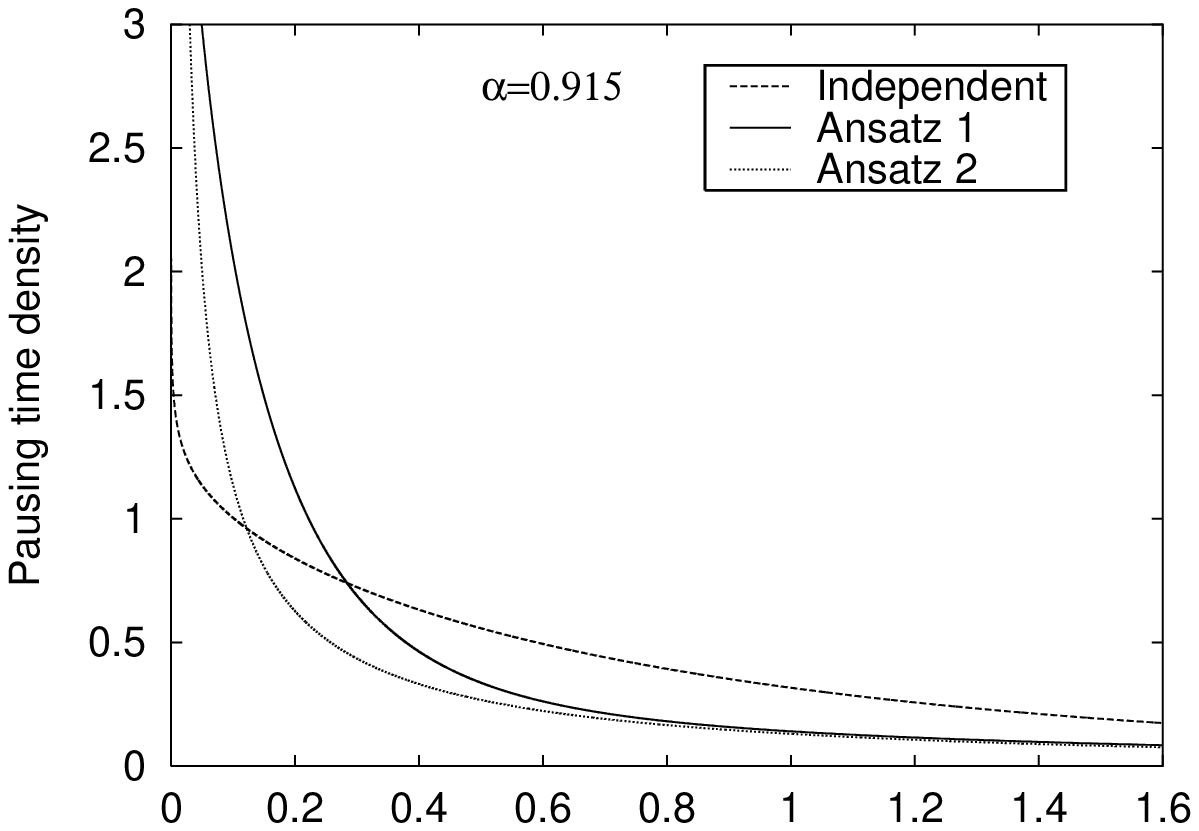,scale=0.52}&\epsfig{file=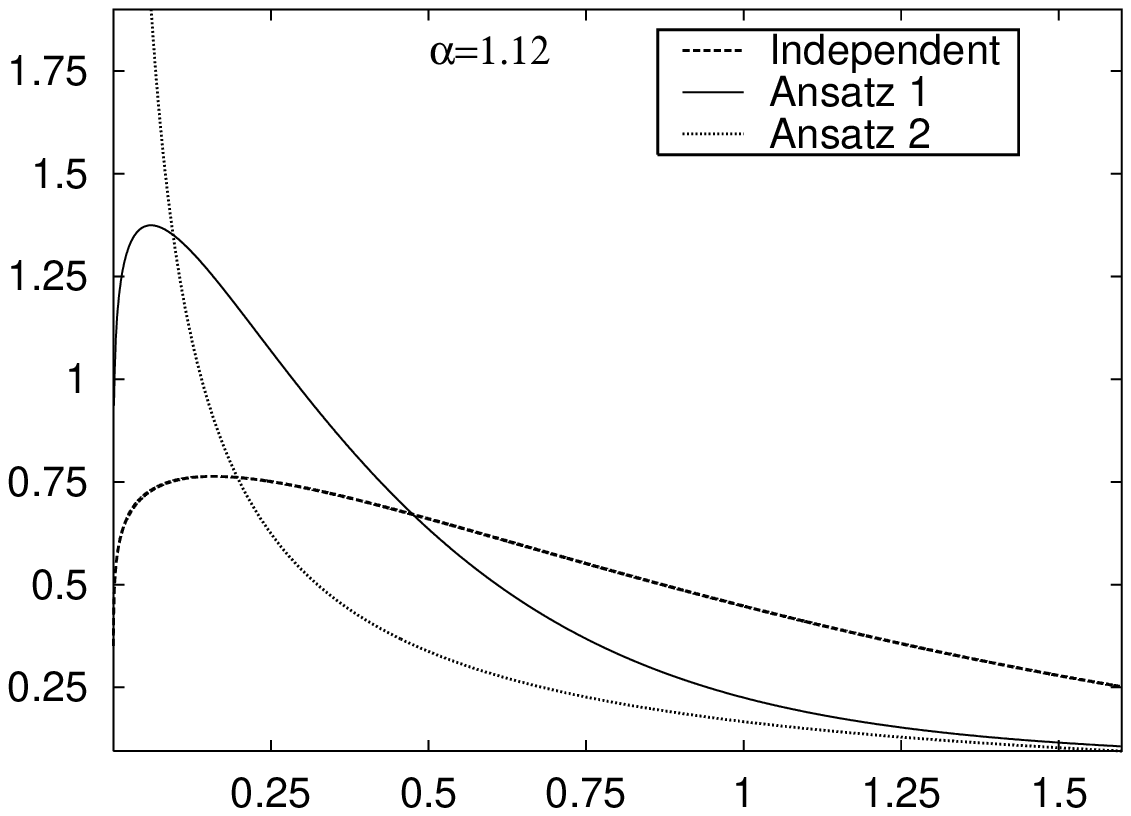,scale=0.52}\\
\epsfig{file=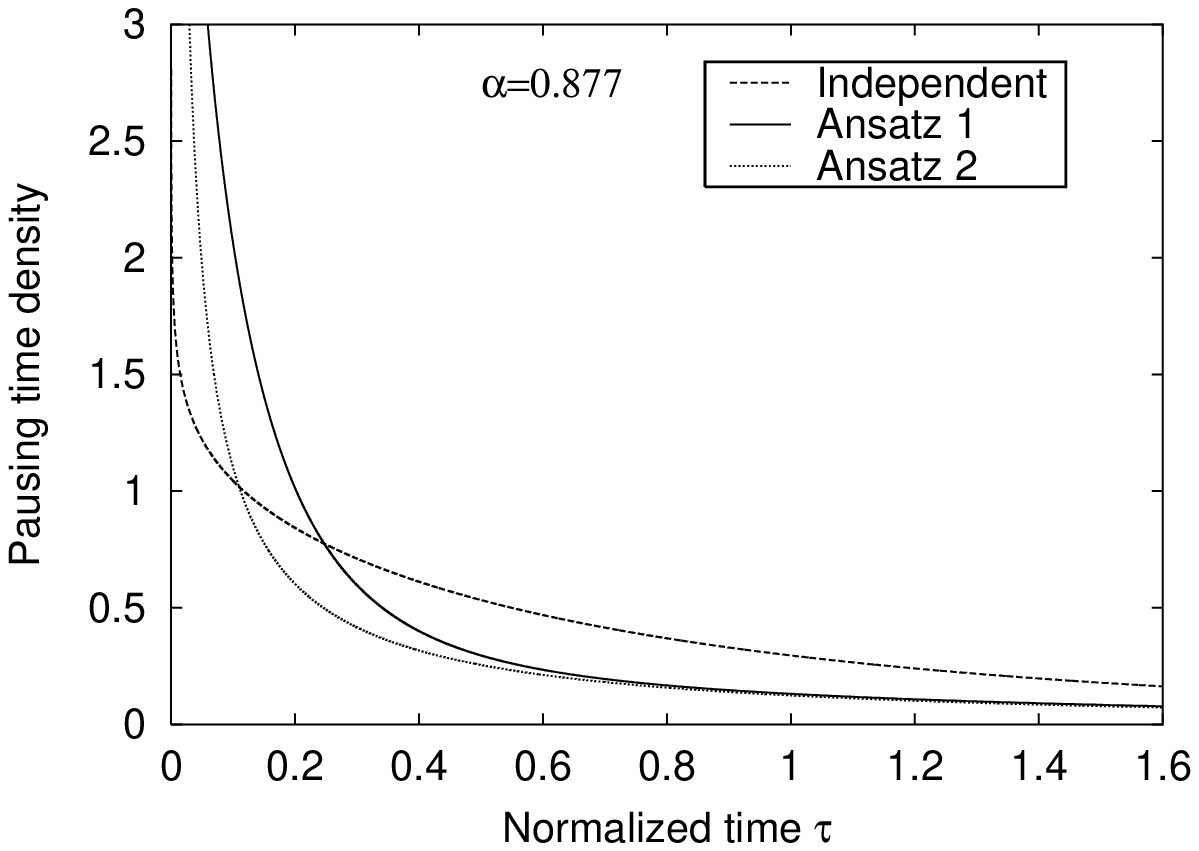,scale=0.52}&\epsfig{file=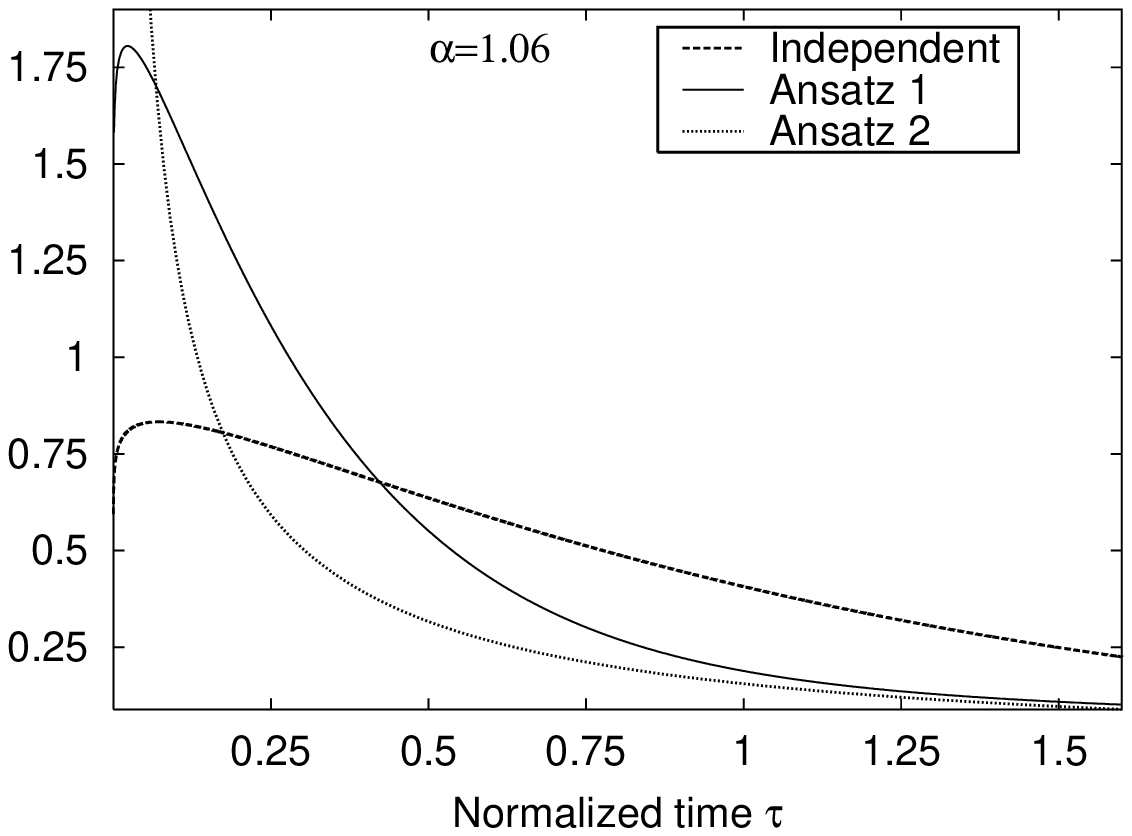,scale=0.52}
\end{tabular}
\caption{The inverse Laplace transform of the pausing time distribution~~(\ref{d}),~(\ref{hatpsipower}) and~(\ref{aaa}) with $\tau$ defined by Eq.~(\ref{tau}). The values of alpha chosen are the ones obtained from IBM ($\alpha=0.915$), 3M ($\alpha=0.877$), Gillette ($\alpha=1.12$) and Dupont ($\alpha=1.06$) data (see Table~\ref{table}).}
\label{invlap1}
\end{figure}

To this point we have determined the return density $p(x,t)$ when the densities $\psi(t)$ and $h(x)$ are known. However, from the CTRW formalism, it is possible to infer the mechanism of price formation at a microscopic level. This requires one to obtain the statistics of high frequency data represented by the pausing-time density $\psi(t)$ and the return increment density $h(x)$ once we know the variance  $\langle X^2(t)\rangle$ and the return pdf $p(x,t)$. We call this ``the inverse problem". Obviously this problem will require the knowledge of the complete pdf $p(x,t)$ for all times $t$, something that is beyond reach, since, in practice, one has access, for instance, to daily data from which one is able to guess the pdf for the daily return $p(x,t)$, where $t=1$ day. In such a case we do not have an analytical expression for $p(x,t)$ permitting us to get the microscopic densities $\psi(t)$ and $h(x)$ exactly. Nevertheless with the formalism presented below we will be able to guess which forms of $\psi(t)$ and $h(x)$ may be consistent with the observed $p(x,t)$. 

Let us first discuss the problem of obtaining the microscopic jump density once we know the return pdf $p(x,t)$. In effect, setting $s=0$ in Eq.~(\ref{hatp}) and taking into account that $\hat{\rho}(\omega,0)=\hat{h}(\omega)$ and $\hat{p}_0(\omega,0)=\langle T\rangle$, we have 
$$
\hat{p}(\omega,0)=\frac{\langle T\rangle}{1-\hat{h}(\omega)}
$$
so that 
\begin{equation}
\hat{h}(\omega)=1-\frac{\langle T\rangle}{\hat{p}(\omega,0)}.
\label{a}
\end{equation}
Since
$$
\hat{p}(\omega,0)=\int_0^\infty \hat{p}(\omega,t) dt,
$$
we see that Eq.~(\ref{a}) allows us to get the characteristic function corresponding to the jump density $h(x)$ if we know the entire return pdf $p(x,t)$ for all $t\geq 0$. Therefore, Eq.~(\ref{a}) is very difficult to implement because, as mentioned, we know at most a numerical series for the value of the propagator instead of an analytical expression of $p(x,t)$. In these cases, Eq.~(\ref{a}) is useless for evaluating $h(x)$. However, Eq.~(\ref{a}) can be useful for testing any statistical hypothesis and evaluating statistical parameters for the form of $h(x)$. Incidentally, we note that Eq.~(\ref{a}) is valid independent of the existing correlations between jumps and pausing times, that is, regardless of the form of the joint density $\rho(x,t)$. 

Unfortunately, the inverse problem for obtaining the pausing time density $\psi(t)$ from the return variance $\langle X^2(t)\rangle$ requires the knowledge of the form of $\rho(x,t)$. Thus, for the independent model in which $\rho(x,t)=h(x)\psi(t)$ the volatility is given by Eq.~(\ref{volatilityindependent}), for which we get
\begin{equation}
\hat{\psi}(s)=\frac{s\hat{m}_2(s)}{\mu_2+s\hat{m}_2(s)}.
\label{b}
\end{equation}
As an illustration of this procedure let us see for which density $\psi(t)$ corresponds to an anomalous diffusion volatility of the form
\begin{equation}
\langle X^2(t)\rangle=kt^\alpha \qquad (k,\alpha>0)
\label{c}
\end{equation}
In this case, 
$$
\hat{m}_2(s)=k\Gamma(1+\alpha)s^{-1-\alpha}
$$ 
where $\Gamma(x)$ is the Gamma function. From Eq.~(\ref{b}), we get
\begin{equation}
\hat{\psi}(s)=\frac{1}{1+\mu_2s^\alpha/k\Gamma(1+\alpha)}.
\label{d}
\end{equation}
We can invert numerically this Laplace transform. To this end we use an algorithm proposed by Stehfest (1970) that is the simplest method for numerically inverting Laplace transforms in the literature. We show in 
Fig.~\ref{invlap1} the pausing time distribution for several values of $\alpha$ in terms of a normalized time $\tau$ defined by
\begin{equation}
\tau=[k\Gamma(1+\alpha)/\mu_2]^{1/\alpha} t.
\label{tau}
\end{equation}

One can also obtain, by using Tauberian theorems (Handelsman and Lew 1974), the asymptotic behavior of $\psi(t)$ as $t\rightarrow\infty$ and also the short time behavior $t\rightarrow 0$. It is worth to mention that Kotulski gives an alternative method based on renewal theory and limit theorems for random sum of jumps. In any case, following Tauberian theorems, we see from Eq.~(\ref{d}) that for small $s$,
$$
\hat{\psi}(s)\simeq 1-\frac{\mu_2}{k\Gamma(1+\alpha)}\ s^\alpha \qquad \left(s^\alpha\ll k\Gamma(1+\alpha)/\mu_2\right),
$$   
and the Tauberian theorems imply that 
\begin{equation}
\psi(t)\simeq \frac{\mu_2\sin(\pi\alpha)}{k\pi}\ t^{-1-\alpha}\qquad \left(kt^{\alpha}\gg \mu_2/\Gamma(1+\alpha)\right).
\label{e}
\end{equation}
On the other hand, for long values of $s$, we have
$$
\hat{\psi}(s)\simeq\frac{k\Gamma(1+\alpha)}{\mu_2}\ s^{-\alpha} \qquad \left( s^\alpha\gg k\Gamma(1+\alpha)/\mu_2\right).
$$ 
Hence,
\begin{equation}
\psi(t)\simeq\frac{\alpha k}{\mu_2}\ t^{-1+\alpha} \qquad \left(kt^{\alpha}\ll \mu_2/\Gamma(1+\alpha)\right).
\label{f}
\end{equation}
We thus see that a power-law volatility leads to a power-law pausing density at short and long times. 

Let us treat now the cases in which there exists a correlation between pausing times and return increments. When the ansatz in Eq. (\ref{fform}) is valid, the formulation of the problem given by Eq. (\ref{formalsolution}) provides a relation between the Fourier and Laplace transforms of the daily return density $p(x,t)$ and the high frequency densities $\psi(t)$ and $h(x)$. A second relation is required to determine the microscopic densities. This is supplied by the volatility. In effect, we start from Eq. (\ref{volatility}), which, on defining $\hat{\phi}(s)\equiv s\hat{\psi}(s)$, can be written in the form
$$
\mu_2\frac{d\hat{\phi}(s)}{ds}+\hat{m}_2(s)\hat{\phi}(s)=s\hat{m}_2(s).
$$
The solution to this equation with the initial condition $\hat{\phi}(0)=0$ yields
\begin{equation}
\hat{\psi}(s)=
\frac{1}{\mu_2 s}\int_0^s\xi\hat{m}_2(\xi)
\exp\left\{-\frac{1}{\mu_2}\int_\xi^s\hat{m}_2(\xi')d\xi'\right\} d\xi.
\label{hatpsi}
\end{equation}
Equation (\ref{hatpsi}) yields an expression for the Laplace transform of the transactions pausing-time density once we know the volatility. 

In the case of a linear variance of the form 
\begin{equation}
\langle X^2(t)\rangle=\sigma^2t,
\label{volbm}
\end{equation}
for which $\hat{m}_2(s)=\sigma^2/s^2$, we see from Eq.~(\ref{hatpsi}) that 
$$
\hat{\psi}(s)=-\frac{D}{s}e^{D/s}Ei(-D/s),
$$ 
where $D=\sigma^2/\mu_2$, and $Ei(z)$ is the exponential integral. The Laplace inversion is (Roberts and Kaufman 1966)
\begin{equation}
\psi(t)=2DK_0(2\sqrt{Dt}),
\label{psibmass(a)}
\end{equation}
where $K_0(x)$ is a modified Bessel function of the second kind (Erdelyi 1954). The asymptotic form of $\psi(t)$ is 
(Magnus et al.)
\begin{equation}
\psi(t)\simeq\left(\frac{\pi^2 D^3}{t}\right)^{1/4}e^{-2\sqrt{Dt}}\qquad (Dt\gg 1).
\label{psibmass}
\end{equation}
Therefore, a linear variance implies that $\ln[\psi(t)]$ falls off assyptotically as $-\sqrt{4Dt}$.

Another example is provided by the anomalous diffusion volatility given in Eq.~(\ref{c}). Now  from Eq. (\ref{hatpsi}) we get
\begin{equation}
\hat{\psi}(s)=\frac{[k\Gamma(\alpha)]^{1/\alpha}}{\mu_2^{1/\alpha}s}U[1/\alpha,1/\alpha,k\Gamma(\alpha)/\mu_2 s^\alpha],
\label{hatpsipower}
\end{equation}
where $U(a,c,x)$ is a Kummer function (Magnus et al.). This function can be inverted numerically with results shown in Fig.~\ref{invlap1}. As before  we use Tauberian theorems to obtain the asymptotic behavior of $\psi(t)$. We have shown in Appendix B that 
\begin{equation}
\psi(t)\simeq\frac{Ak}{\mu_2}\ t^{\alpha-1} \qquad (kt^{\alpha}\ll\mu_2),
\label{asympsi1}
\end{equation}
and
\begin{equation}
\psi(t)\simeq\frac{\mu_2\sin(\pi\alpha)}{k\pi}\ t^{-1-\alpha} \qquad (kt^{\alpha}\gg\mu_2),
\label{asympsi2}
\end{equation}
where $A$ is a constant given by Eq.~(\ref{a2}). Note that these results agree with Eqs.~(\ref{e}) and~(\ref{f}).

Let us briefly comment on the inverse problem under the assumption that pausing times and return jumps are correlated as in ansatz~(\ref{hatrho2}). In this case from Eq.~(\ref{volatilitygeneral}) we have
\begin{equation}
\hat{m}_2(s)=\frac{\mu_2 \hat{\psi}^3(s)}{s\left[1-\hat{\psi}(s)\right]}.
\label{aa}
\end{equation}
Consequently, if we know the return moment $\langle X^2(t)\rangle$ and hence its Laplace transform $\hat{m}_2(s)$, the Laplace transform of the microscopic pausing-time density $\psi(t)$ obeys the cubic algebraic equation
\begin{equation}
\mu_2\hat{\psi}^3(s)+s\hat{m}_2(s)\hat{\psi}(s)-s\hat{m}_2(s)=0, 
\label{bb}
\end{equation}
whose solution is
\begin{equation}
\hat{\psi}(s)=\hat{\beta}(s)^{1/3}
\left\{\left[1+\sqrt{1+8\hat{\beta}(s)/27}\right]^{1/3}+
\left[1-\sqrt{1+8\hat{\beta}(s)/27}\right]^{1/3}\right\},
\label{aaa}
\end{equation}
where $\hat{\beta}(s)\equiv s\hat{m}_2(s)/2\mu_2$. The numerical inversion of this equation when $\langle X^2(t)\rangle=kt^\alpha$ is shown in Fig.~\ref{invlap1}. Again we can proceed as before and perform an asymptotic analysis based on the use of Tauberian theorems. We present the results of these calculations without going into details. We have
\begin{equation}
\psi(t)\simeq \frac{k\Gamma(1+\alpha)}{\mu_2\Gamma(\alpha/3)}\ t^{-1+\alpha/3}\qquad \left(kt^{\alpha}\ll \mu_2\right),
\end{equation}
and
\begin{equation}
\psi(t)\simeq\frac{\mu_2\sin(\pi\alpha)}{k\pi}\ t^{-1-\alpha} \qquad (kt^{\alpha}\gg\mu_2),
\label{bbb}
\end{equation}
which is similar to Eqs.~(\ref{e})--(\ref{f}) and~(\ref{asympsi1})--(\ref{asympsi2}). We therefore see that the appearance of asymptotic power-law pausing densities seems to be independent of the existence of correlations between jumps and pausing times. We also observe the asymptotic distribution at long times has the same decay exponent for dependent and independent models.

\begin{figure}
\begin{tabular}{c}
\epsfig{file=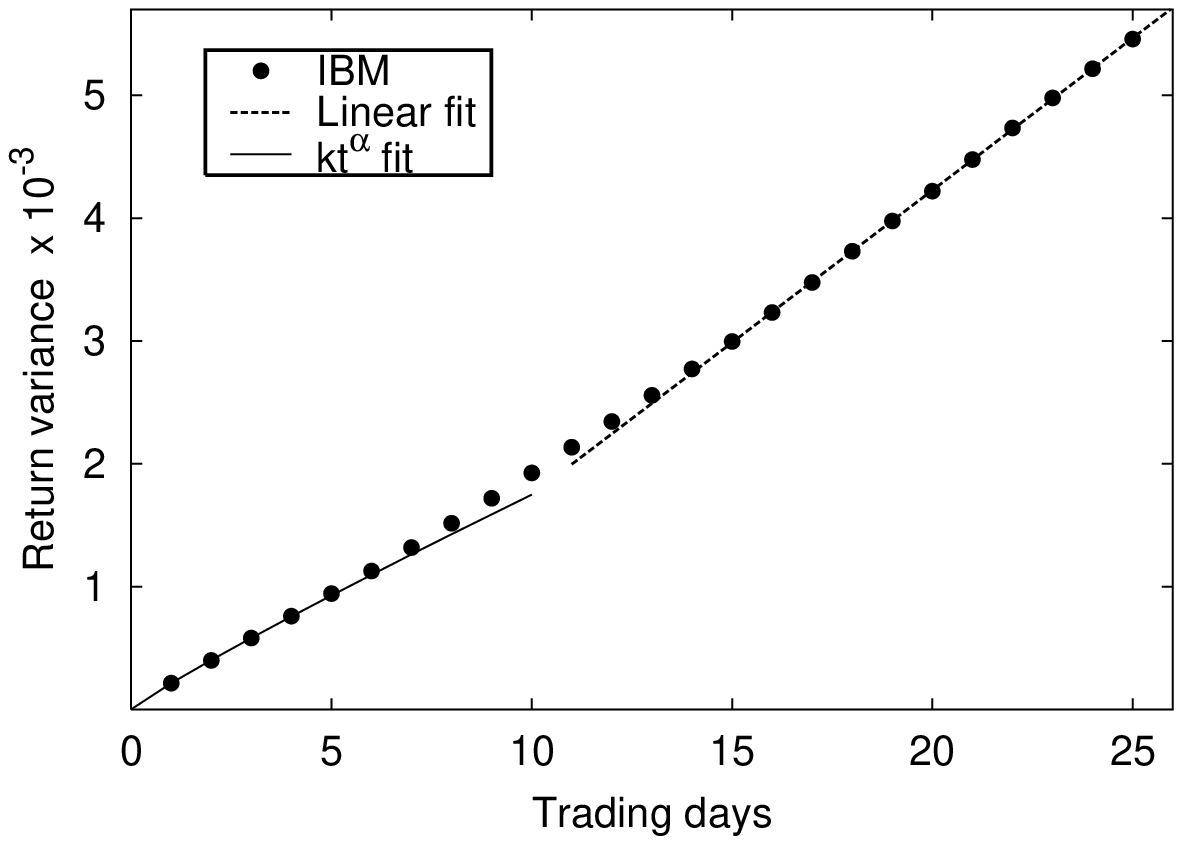,scale=0.55}\epsfig{file=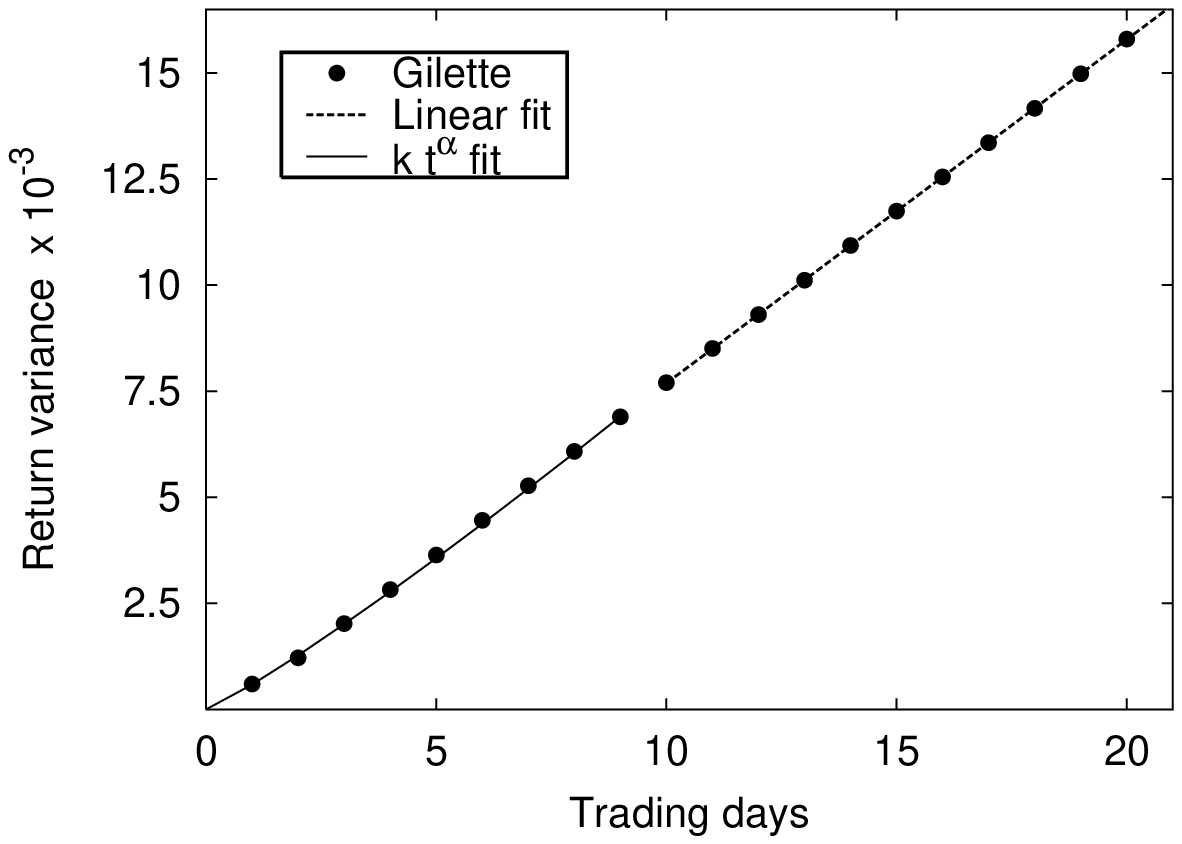,scale=0.55}\\
\epsfig{file=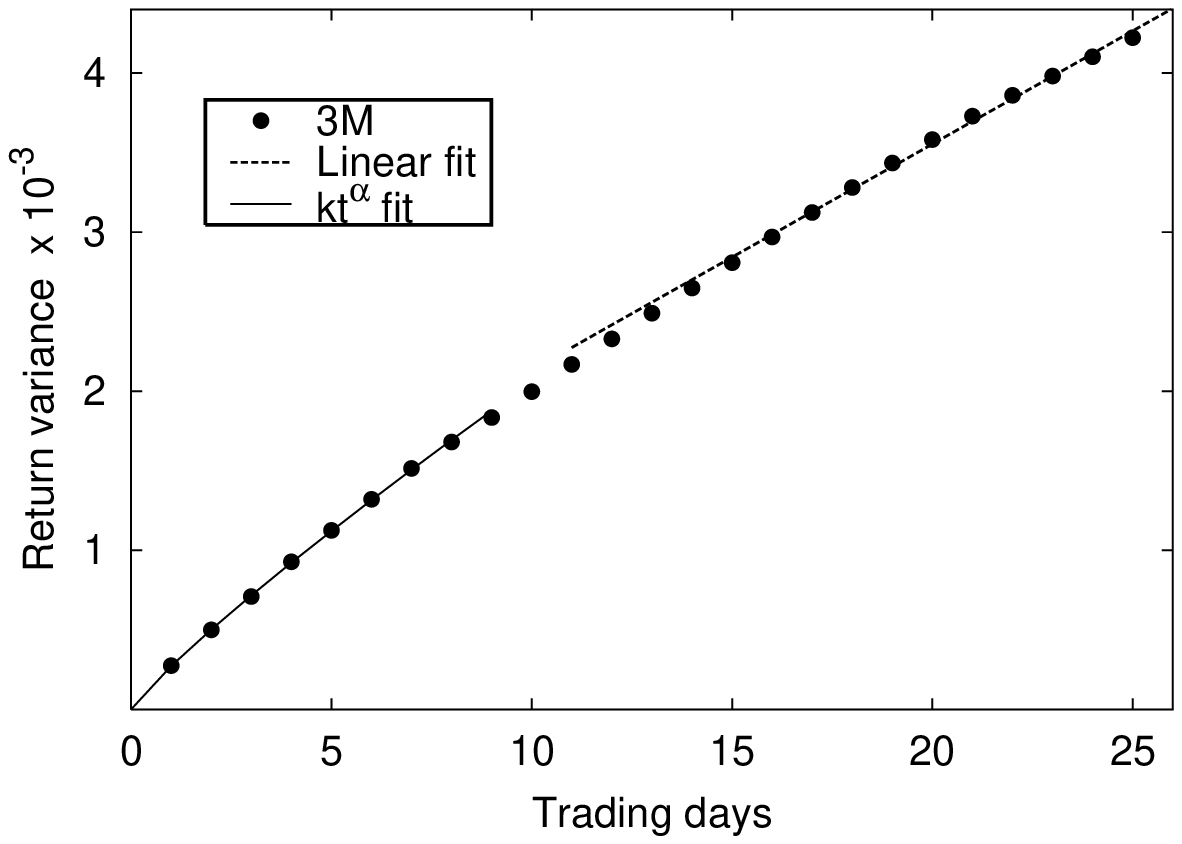,scale=0.55}\epsfig{file=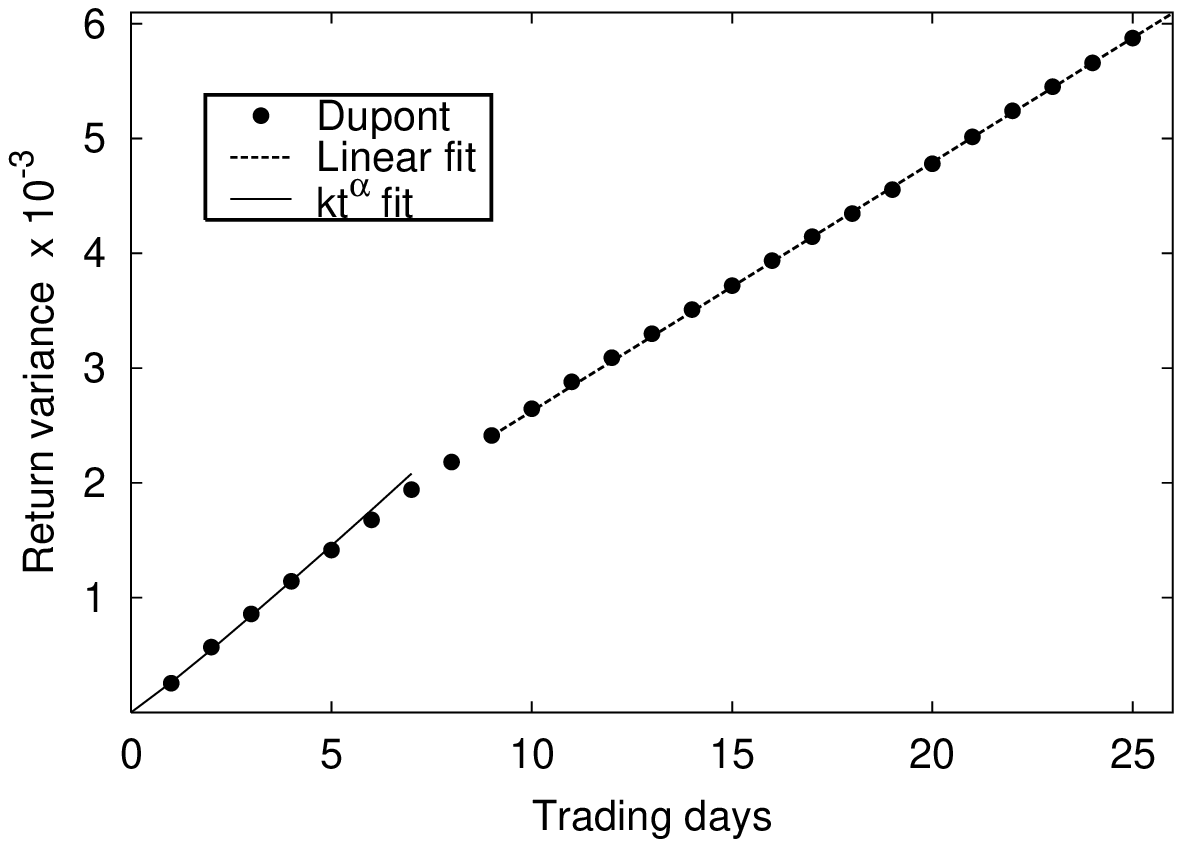,scale=0.55}
\end{tabular}
\caption{The return variance dependence with time for IBM, Gillette, 3M, and Dupont.} 
\label{figvar}
\end{figure}

\begin{table}
\caption{Summary of the linear and polynomial regressions for data from IBM, Gillette, 3M and Dupont. The details of the linear and polynomial regressions are shown specifying the range within each fit is performed and the values with their errors obtained from each regression. Some parameters are given in days (d).}
\bigskip
\begin{center}
\begin{tabular}{lcccc}
\hline
\hline
& IBM & Gillette & 3M & Dupont\\
\hline
Period &  1968-2002 & 1980-2002 & 1980-2003 & 1980-2003\\
Points & 8,784 & 5,628 & 5,842 & 5,801 \\
\\
\multicolumn{2}{l}{Linear Regression: $\langle X^2(t)\rangle=At+B$}\\
\hline
Range in days& 15--25& 11--20 & 15--25& 15--25\\
$(A\pm\delta A)\times 10^{4} \mbox{d}^{-1}$ & $2.5\pm 0.7$&$8.104\pm 0.005$&$1.42\pm0.03$&$2.17\pm 0.01$\\
$(B\pm\delta B)\times 10^{4}$ & $7.30\pm 0.14$&$4.15\pm 0.07$&$7.1\pm 0.5$&$4.6\pm 0.2$\\
\\
\multicolumn{2}{l}{Polynomial Regression: $\langle X^2(t)\rangle=kt^\alpha$}\\
\hline
Range in days& 1--4& 1--4 & 1--5& 1--5\\
$(k\pm\delta k)\times 10^{4} \ \mbox{d}^{-1/\alpha}$ & $2.129\pm 0.010$&$5.8\pm 0.2$&$2.73\pm 0.02$&$2.61\pm0.07$\\
$\alpha\pm\delta \alpha$ & $0.915\pm 0.005$&$1.12\pm 0.04$&$0.877\pm 0.006$&$1.06\pm 0.03$\\
\hline \hline
\bigskip
\end{tabular}
\end{center}
\label{table}
\end{table}

\begin{figure}
\begin{center}
\epsfig{file=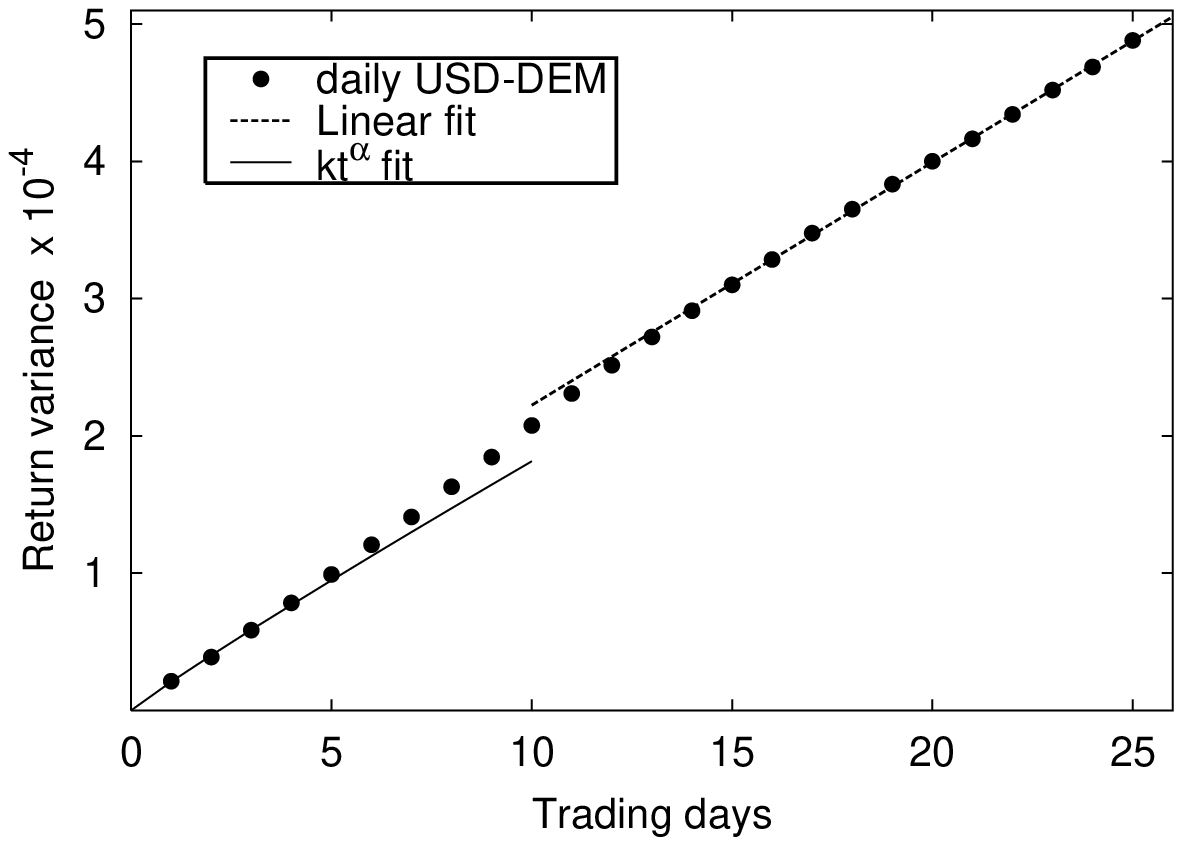}
\end{center}
\caption{The U.S. dollar/Deutsche mark daily return variance ranging from 1 day to 25 days return changes. We have filtered the tick-by-tick futures data to provide the daily return changes from January, 1993, to December, 1997, 1,035 days. Again the short time scale is fitted assuming that the variance has the time dependence $kt^{\alpha}$. The polynomial regression is done with nine data points (from 1 day until 9 days) and yields  $k=(2.09\pm 0.05)\times 10^{-5}(\mbox{days})^{1/\alpha}$ and $\alpha=0.94\pm 0.03$. The long time fit assumes a linear behavior, that is, $At+B$. The linear regression takes eleven  data points (from 15 to 25 days) and yields  $A=(1.42\pm0.03)\times 10^{-4}(\mbox{days})^{-1}$ and $B=(7.1\pm 0.5)\times 10^{-4}$.} 
\label{fig10}
\end{figure}

\subsection{A market application}\label{marketapplication}

We finally present an example involving real data. We base our discussion on four sets of data points corresponding to daily opening and closing prices for shares of IBM, Gillette, Dupont and 3M. These were chosen because a considerable amount of data was available for them. Thus data for IBM consists of daily prices from Jan. 2, 1968 to Dec. 31, 2002 with a total of 8,784 data points. The data for Gillette corresponds to daily prices from July 27, 1980 to Dec. 31, 2002 with 5,628 data points. Analogously the data for Dupont and 3M consist of  daily prices from July 28, 1980 to Oct. 17, 2003 with  respectively a total of 5,801 and 5,842 points. For all these firms we have tried to deduce the form of the microscopic pausing time density, $\psi(t)$, from a knowledge of low-frequency volatilities. We first plot $\langle X^2(t)\rangle$ for $t\ge 1$ day. 
This is shown in Fig.~\ref{figvar}. In the case of all of these companies we see a slightly but clear anomalous diffusive behavior for few days and a regular diffusive behavior at long times:
\begin{equation}
\langle X^2(t)\rangle\sim t^\alpha\quad(t\rightarrow 0)\qquad\mbox{and}\qquad
\langle X^2(t)\rangle\sim t\quad(t\rightarrow\infty),
\label{ibmvol}
\end{equation}
where $\alpha=0.915\pm 0.005$ (IBM), $\alpha=1.12\pm 0.04$ (Gillette), $\alpha=1.06\pm 0.03$ (Dupont) and $\alpha=0.877\pm 0.006$ (3M). We therefore see that the anomalous diffusive behavior in the data is subdiffusive for IBM and 3M and superdiffusive  for Gillette and Dupont. We summarize all the estimations of these fits in Table~\ref{table}.

\begin{figure}
\begin{center}
\epsfig{file=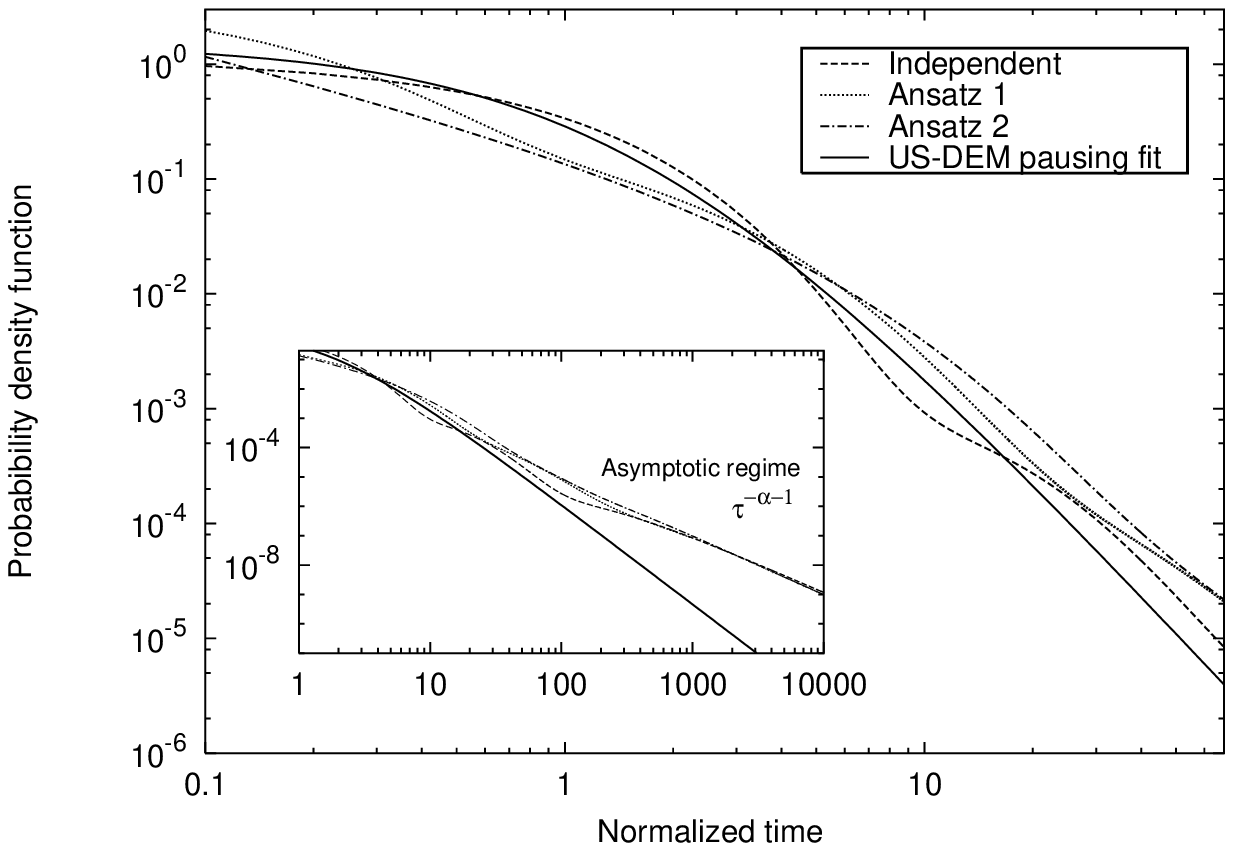}
\end{center}
\caption{The pausing time distributions for the U.S. dollar/Deutsche mark. We compare the pausing time distributions obtained from the Laplace inversion of Eqs.~(\ref{d}),~(\ref{hatpsipower}) and~(\ref{aaa}) with the empirical pausing time distribution~(\ref{psifit}). The results produced by the three ansatzes are polotted with the value of $\alpha$ obtained from the fit of fewer days in Fig.~\ref{fig10} with $\tau$ given by Eq.~(\ref{tau}). The empirical fit takes $\tau=t/\langle T\rangle$. Note that both normalized times are quite similar. The inset shows in a logscale how the three models collapse in the same power law when $\tau>1,000$.} 
\label{fig11}
\end{figure}

What can we say about the microscopic density $\psi(t)$? From the analysis performed above we see that, in all cases, $\psi(t)$ dacays exponentially at long times. For intermediate times, times not too long neither too short, for instance $t\geq (\mu_2/k)^{1/\alpha}$ (see Eq. (\ref{c})), we cannot apply either Tauberian or Abelian theorems. In these cases, as we have done above (see Fig.~\ref{invlap1}), we must numerically invert $\hat{\psi}(s)$ to obtain $\psi(t)$.

Let us remark that a good empirical test of the formalism used for the inverse problem is to estimate the waiting time density. Unfortunately, we could not get enough data to perform such a test for the four stocks studied above. However, we can take the U.S. dollar/Deutsche mark high frequency data and filter them in order to get daily return changes. We thus observe the time-dependence of the variance provided by this daily data and perform a  fit to the form $kt^\alpha$. Figure~\ref{fig10} plots the resulting variance and its fit. 

Once we have estimated the $\alpha$, we can numerically invert $\hat{\psi}(s)$ and finally we can compare the pausing time distribution provided by the inverse method with the one directly observed from high-frequency data. We compare them in Fig.~\ref{fig11} showing a good agreement in the region of intermediate times for which $kt^{\alpha}\geq\mu_2$. 

If we take a closer look at the three pausing distributions provided by the independent model and by the dependent models exemplified by ansatzes 1 and 2, we observe that both the independent model and ansatz 1 describe the U.S. dollar/Deutsche mark data much better than does ansatz 2. This is explained by considering the quantity $r$ defined in Eq. (\ref{r}) that quantifies the correlation between waiting times and jumps. We have seen that for the independent model $r=0$, ansatz 1 gives $r=1$, and ansatz 2, $r=2$. On the other hand, the empirical correlation of the US dollar/Deutsche mark data is (see Section~\ref{rp}) $r=0.53$, and intermediate value between those of the independent model and ansatz 1, while the correlation given by ansatz 2 is higher. All of this seems to indicate that the actual Foreign Exchange (FX) market should be described with an intermediate model between the independent model and ansatz 1 while the model based on ansatz 2 is too correlated. This is exactly what is observed in Fig.~\ref{fig11} since the curves corresponding to the independent model and ansatz 1  oscillate very closely around the empirical distribution from $\tau=0.1$ to $\tau=20$ while the curve corresponding to ansatz 2 clearly deviates from the empirical result. 

We can quantify this by performing the chi-square test on these fits. The results, based on Fig.~\ref{fig11}, are the following. The value of chi-square statistic for the independent model is 6.5 times higher than that of the fitted distribution given by Eq.~(\ref{psifit}). For ansatz 1 the ratio is 39.2, and for the ansatz 2 the value is 60.7 times higher than that of the pausing fit. This is quite consistent with the assertions in the previous paragraph.

We finally observe from the inset of Fig.~\ref{fig11} that for large times, $\tau\sim 1,000$, the three models converge to the same power law decay as was anticipated by 
Eqs.~(\ref{e}),~(\ref{asympsi2}), and~(\ref{bbb}) with $\alpha+1=1.94$, which differs from the behavior of the empirical pausing-time density shown in Fig.~\ref{fig11}. This is not surprising since this time scale is outside the region of validity of the empirical distribution (see Fig.~\ref{fig2} and Eq.~(\ref{psifit})). Indeed, if we assume that $t\sim\tau\langle T\rangle$, then  $\tau=1,000$ approximately corresponds to seven hours of trading, an interval of time exceeding the trading day. 

\section{Conclusions\label{conclusions}}

We have described some uses of CTRW formalism in the context of financial markets. The CTRW formalism provides insight that is able to relate not only the market microstructure activity with the probability distribution of intraday prices, but also with the distributions of daily, weekly, or longer-time prices. Techniques based on the CTRW also can be applied to  inverse problems in which it is possible to analyze the microstructure of financial processes, represented by their transaction statistics, based only on low-frequency (daily, weekly, etc.) data. In this way, we can have some knowledge of the market microstructure by taking the information related to low frequency price changes. 

The formalism, in its direct version (i.e., from high-frequency data to lower frequency statistics), depends on estimates of the pausing-time density, $\psi(t)$, and the jump density, $h(x)$. A first and simpler approach consists in assuming that these densities are independent. However, in many practical situations one certainly expects some degree of correlation, in the sense that large return increments are infrequent. We have proposed two possible ways to model this correlation. We have described some general features of the formalism that hold for any model of correlation between sojourn times and return jumps. Thus (i) the variance grows linearly with time as $t$ increases, and (ii) the tails of the return distribution $p(x,t)$ have the same appearance than those of the jump density $h(x)$. We have also applied the direct formalism to data provided by the U.S. dollar/Deutsche mark future exchange rate. High frequency data indicate that $\psi(t)$ and $h(x)$ are well described by power-law densities. We have therefore shown that the volatility has a diffusion-like behavior at long times and that the tails of the return distribution, $p(x,t)$, follow a power law with the same exponent as that of $h(x)$. 

We have applied the inverse formalism to daily data provided by four companies: IBM, Gillette, Dupont and 3M. We have focused on the volatility and its behavior as a function of the time lag. The data for all companies exhibit a slight but clearly anomalous diffusive  behavior at short times since the variance can be fitted with a polynomial function growing as $t^\alpha$ in the short time regime (few days), and as was expected, the variance describes an ordinary diffusion-like behavior for return changes of a week or more, that is, a linear dependence on time for large time lags. All of this implies that the shortest-day behavior of the unknown pausing-time density $\psi(t)$ follows a power law consistent with high frequency data for the US-Deutsche mark case and that for larger times the pausing-time density decays exponentially.

Finally, these results seem to indicate the suitability of the CTRW framework to study other complex issues in finance. We can mention that the present formalism can be useful in the study of problems related to large fluctuations and extreme values central to risk control and also to exotic derivative pricing. All of these are under present investigation.

\ack

This work has been supported in part by Direcci\'on General de Investigaci\'on under contract No. BFM2003-04574 and by Generalitat de Catalunya under contract No. 2000 SGR-00023. We acknowledge J. M. Porr\`a and Gaesco Bolsa S.V. y B. for providing daily data.

\appendix

\section{Asymptotic results}\label{apa}

Let us prove the results given in Section~\ref{sec:asymp}. Since $\Delta X$ and $T$ are random variables representing respectively returns increments and sojourn times, we can write $\hat{\rho}(\omega,s)$ in the form (see Eq. (\ref{rhodef}))
$$
\hat{\rho}(\omega,s)=\langle\exp\{i\omega\Delta X-sT\}\rangle. 
$$ 
Expanding this average around $s=0$, we get
\begin{equation}
\hat{\rho}(\omega,s)\simeq\tilde{h}(\omega)-s\langle Te^{i\omega\Delta X}\rangle
\qquad \left(s\ll \langle T\rangle^{-1}\right),
\label{asymrho}
\end{equation}
where 
$$
\tilde{h}(\omega)=\langle e^{i\omega\Delta X}\rangle.
$$
On the other hand, if we assume that the sojourn-time density $\psi(t)$ has a finite first moment, then 
\begin{equation}
\hat{\psi}(s)\simeq 1-s\langle T\rangle\qquad \left(s\ll \langle T\rangle^{-1}\right),
\label{hatpsi0}
\end{equation}
where $\langle T\rangle$ is the mean sojourn time. Therefore, from Eq.~(\ref{hatp0}) we have 
\begin{equation}
\hat{p}_0(\omega,s)\simeq\langle T\rangle\qquad \left(s\ll \langle T\rangle^{-1}\right).
\label{asymp0}
\end{equation}
Substituting Eqs. (\ref{asymrho}) and (\ref{asymp0}) into Eq.~(\ref{hatp}) yields
\begin{equation}
\hat{p}(\omega,s)\simeq\frac{\langle T\rangle}
{1-\tilde{h}(\omega)+s\langle Te^{i\omega\Delta X}\rangle} \qquad\left(s\ll \langle T\rangle^{-1}\right).
\label{asymhatp}
\end{equation}

By virtue of Tauberian theorems (Handelsman and Lew 1974), the asymptotic behavior of the characteristic function $\tilde{p}(\omega,t)$ as $t\rightarrow\infty$ will be given by the inverse Laplace transform of Eq. (\ref{asymhatp}). The resulting inverse reads
\begin{equation}
\tilde{p}(\omega,t)\simeq\frac{\langle T\rangle}
{\langle Te^{i\omega\Delta X}\rangle}
\exp\left\{-\frac{[1-\tilde{h}(\omega)]t}{\langle Te^{i\omega\Delta X}\rangle}\right\}
\qquad \left( t\gg \langle T\rangle\right).
\label{asymp}
\end{equation}
We emphasize that the asymptotic expression in Eq. (\ref{asymp}) is totally general regardless of the correlation between sojourn times $T$ and return increments $\Delta X$, requiring only that the mean sojourn time $\langle T\rangle$ be finite. Obviously for a  usable form of this asymptotic expression we will need to know the correlation between $T$ and $\Delta X$. In other words, we will have to specify a form for $\rho(x,t)$ since the correlation appearing in Eq.~(\ref{asymp}) is related to $\hat{\rho}$ by
$$
\langle Te^{i\omega\Delta X}\rangle=
-\left.\frac{\partial}{\partial s}\hat{\rho}(\omega,s)\right|_{s=0}.
$$
Thus, for instance, making use of ansatz in Eq. (\ref{fform}), we have
$$
\langle Te^{i\omega\Delta X}\rangle=\langle T\rangle\tilde{h}^2(\omega)
$$
so that
\begin{equation}
\tilde{p}(\omega,t)\simeq\frac{1}{\tilde{h}^2(\omega)}
\exp\left\{-\frac{[1-\tilde{h}(\omega)]t}{\langle T\rangle\tilde{h}^2(\omega)}\right\}
\qquad\left(t\gg \langle T\rangle\right).
\label{asympansatz}
\end{equation}

Now assume that $h(x)$ has a finite second moment $\mu_2$ so that  $\tilde{h}(\omega)\simeq 1-\mu_2\omega^2/2$ as $\mu_2\omega^2\ll 1$. In this case, for values of $\omega$ sufficiently small, Eq. (\ref{asymp}) yields the Gaussian density
$$
\tilde{p}(\omega,t)\simeq e^{-\mu_2t\omega^2/2\langle T\rangle},
$$
which is equivalent to the result given by the central limit theorem.

On the other hand, for a long-tailed jump density for which $\tilde{h}(\omega)\simeq 1-k|\omega|^{\alpha}$ for small $\omega$ (note that if $0<\alpha<2$, $h(x)$ has infinite variance) and assuming that for 
$\omega$ small $\langle Te^{i\omega\Delta X}\rangle\simeq\langle T\rangle$, we obtain from Eq. (\ref{asymp}) the L\'{e}vy distribution 
$$
\tilde{p}(\omega,t)\simeq e^{-k|\omega|^\alpha t/\langle T\rangle}.
$$

We finally show that the tails of the return pdf $p(x,t)$ are the same as that of the jump pdf $h(x)$. Indeed, as is well known the tails of $p(x,t)$ as $|x|\rightarrow~\infty$ are determined by the behavior of its characteristic function as $\omega\rightarrow 0$ (Weiss 1994). Thus suppose that $\omega$ is small so that $\langle Te^{i\omega\Delta X}\rangle\simeq\langle T\rangle$ and $t$ is moderate in the sense that the approximation given in Eq. (\ref{asymp}) is still valid but in such a way that $[1-\tilde{h}(\omega)]t/\langle T\rangle\ll 1$. Then from Eq.~(\ref{asymp}) we see that for small $\omega$,
$$
\tilde{p}(\omega,t)\sim-[1-\tilde{h}(\omega)]t/\langle T\rangle,
$$
which for large values of $|x|$ is equivalent to
$$
p(x,t)\sim\frac{t}{\langle T\rangle}h(x),\qquad\left(t\approx\langle T\rangle\right).
$$
Therefore, we have proved Eq.~(\ref{tailsp}) in item (c) of Section~\ref{sec:asymp} and shown that the tails of $p(x,t)$ are equal to the tails of $h(x)$.

\section{Pausing time density limits}

In order to get the asymptotic behavior of $\psi(t)$ as $t\rightarrow\infty$ and as $t\rightarrow 0$, we first write the Kummer function appearing in Eq. (\ref{hatpsipower}) in terms of the incomplete gamma function: $U(a,a,x)=e^x\Gamma(1-a,x)$ (Magnus et al. 1966). Therefore,
\begin{equation}
\hat{\psi}(s)=\frac{[k\Gamma(\alpha)]^{1/\alpha}}{\mu_2^{1/\alpha} s}e^{k\Gamma(\alpha)/\mu_2 s^\alpha}
\Gamma\left[1-1/\alpha,k\Gamma(\alpha)/\mu_2 s^\alpha\right].
\label{a1}
\end{equation}
Now, using Tauberian theorems (Handelsman and Lew 1974), the behavior as $t\rightarrow 0$ of $\psi(t)$ will be given by the behavior of $\hat{\psi}(s)$ as $s\rightarrow\infty$ while the behavior of $\psi(t)$ as $t\rightarrow\infty$ is determined by $\hat{\psi}(s)$ as $s\rightarrow 0$.

(a) {\it Short-time behavior}. We use the following series representation of the incomplete gamma function:
$$
\Gamma(\beta,x)=e^{-x}x^\beta\sum_{n=0}^{\infty}\frac{L_n^{(\beta)}(x)}{n+1}=
e^{-x}x^\beta\left[\sum_{n=0}^{\infty}\frac{L_n^{(\beta)}(0)}{n+1}+O(x)\right],
$$
where 
$$
\beta=1-1/\alpha,\qquad x=k\Gamma(\alpha)/\mu s^\alpha, 
$$
and $L_n^{(\beta)}(x)$ are the generalized Laguerre polynomials (Magnus et al.). Since $L_n^{(\beta)}(0)=\Gamma(1+\beta+n)/n!\Gamma(1+\beta)$, we have
$$
\Gamma(\beta,x)=e^{-x}\left[Ax^{\beta}+O\left(x^{\beta+1}\right)\right],
$$
where
\begin{equation}
A=\sum_{n=0}^{\infty}\frac{\Gamma(1+\beta+n)}{n!\Gamma(1+\beta)}.
\label{a2}
\end{equation}
Hence
$$
\hat{\psi}(s)=\frac{Ak\Gamma(\alpha)}{\mu_2}\ s^{-\alpha}+O\left(\frac{\mu_2 s^{-2\alpha}}{k^2\Gamma(\alpha)^2}\right),
$$
which implies (Handelsman and Lew)
\begin{equation}
\psi(t)\simeq\frac{Ak}{\mu_2}\ t^{\alpha-1} \qquad \left(kt^{\alpha}\ll{\mu_2}\right).
\label{a3}
\end{equation}

(b) {\it Long-time behavior}. In order to obtain the long-time behavior of the density $\psi(t)$ corresponding to the volatility, we employ the following asymptotic behavior for the gamma function (Magnus et al.):
$$
\Gamma(\beta,x)=e^{-x}x^{\beta-1}\left[1-(1-\beta)\frac{1}{x}+
O\left(\frac{1}{x^2}\right)\right].
$$
Thus
$$
\hat{\psi}(s)=1-\frac{\mu_2 s^\alpha}{\alpha k\Gamma(\alpha)}+
O\left(\frac{\mu_2s^{2\alpha}}{k^2\Gamma(\alpha)^2}\right),
$$
and (Handelsman and Lew)
\begin{equation}
\psi(t)\simeq \frac{\mu_2}{k\Gamma(-\alpha)\Gamma(1+\alpha)}\ t^{-1-\alpha} \qquad \left(kt^{\alpha}\gg\mu_2\right).
\label{a4}
\end{equation}

\end{document}